\documentclass[acmsmall,screen]{acmart}

\newcommand{\rs}{0}
\usepackage{setspace}
\usepackage{times}
\usepackage{soul}
\usepackage{url}
\usepackage[inline]{trackchanges}

\usepackage[utf8]{inputenc}
\usepackage{graphicx}
\usepackage{amsmath}
\usepackage{amsthm}
\usepackage{booktabs}
\usepackage{algorithm}
\usepackage{algorithmic}
\usepackage{array}
\usepackage{makecell}
\usepackage{float}

\usepackage[british]{babel}
\usepackage{hhline}
\DeclareUnicodeCharacter{FFFC}{ }
\usepackage{multirow}
\usepackage{color}
\usepackage{booktabs}
\usepackage{amsmath}
\usepackage{graphicx}
\usepackage{subfigure}
  \usepackage{dsfont}
 \usepackage{enumerate}
 \usepackage{empheq}
 \usepackage{caption} 
 \usepackage{amsthm, amsfonts}
\usepackage{totcount}
\usepackage{endnotes}
\let\svthefootnote\thefootnote
\newcommand\freefootnote[1]{%
  \let\thefootnote\relax%
  \footnotetext{#1}%
  \let\thefootnote\svthefootnote%
}
\AtBeginDocument{%
  \providecommand\BibTeX{{%
    \normalfont B\kern-0.5em{\scshape i\kern-0.25em b}\kern-0.8em\TeX}}}
\setcopyright{acmcopyright}
\copyrightyear{2020}
\acmYear{2020}
\acmDOI{10.1145/1122445.1122456}

\begin{document}

\newcommand{\yell}[1]{{\color{red} \textbf{#1}}}
\newcommand\blfootnote[1]{%
  \begingroup
  \renewcommand\thefootnote{}\footnote{#1}%
  \addtocounter{footnote}{-1}%
  \endgroup
}

\title[Physics-Guided Machine Learning Survey]{Integrating Scientific Knowledge with Machine Learning for Engineering and Environmental Systems}

\author{Jared Willard}
\authornote{Both authors contributed equally to this research.}
\email{willa099@umn.edu}
\orcid{0000-0003-4434-051X}
\affiliation{%
  \institution{University of Minnesota}
  \city{Minneapolis}
  \state{Minnesota}
  \postcode{55455}
}
\author{Xiaowei Jia}
\authornotemark[1]
\email{xiaowei@pitt.edu}
\orcid{0000-0001-8544-5233}
\affiliation{%
  \institution{University of Pittsburgh}
  \city{Pittsburgh}
  \state{Pennsylvania}
  \postcode{15260}
}
\author{Shaoming Xu}
\email{xu000114@umn.edu}
\affiliation{%
  \institution{University of Minnesota}
  \city{Minneapolis}
  \state{Minnesota}
  \postcode{55455}
}
\author{Michael Steinbach}
\email{stei0062@umn.edu}
\affiliation{%
  \institution{University of Minnesota}
  \city{Minneapolis}
  \state{Minnesota}
  \postcode{55455}
}
\author{Vipin Kumar}
\email{kumar001@umn.edu}
\affiliation{%
  \institution{University of Minnesota}
  \city{Minneapolis}
  \state{Minnesota}
  \postcode{55455}
}

\begin{abstract}

There is a growing consensus that solutions to complex science and engineering problems require novel methodologies that are able to integrate traditional physics-based modeling approaches with state-of-the-art machine learning (ML) techniques.  This paper provides a structured overview of such techniques. Application-centric objective areas for which these approaches have been applied are summarized, and then classes of methodologies used to construct physics-guided ML models and hybrid physics-ML frameworks are described. We then provide a taxonomy of these existing techniques, which uncovers knowledge gaps and potential crossovers of methods between disciplines that can serve as ideas for future research.

\end{abstract}
\freefootnote{This work was supported by NSF grant \#1934721 and by DARPA award W911NF-18-1-0027.}

\maketitle

\section{Introduction}
\label{sect:introduction}
Machine learning (ML) models, which have already found tremendous success in commercial applications, are beginning to play an important role in advancing scientific discovery in environmental and engineering domains traditionally dominated by mechanistic (e.g. first principle) models \cite{hsieh2009machine,kutz2017deep,karpatne2017theory,karpatne2018machine,wang2018successful,bergen2019machine,reichstein2019deep,ivezic2019statistics}. The use of ML models is particularly promising in scientific problems involving processes that are not completely understood, or where it is computationally infeasible to run mechanistic models at desired resolutions in space and time. However, the application of even the state-of-the-art black box ML models has often met with limited success in scientific domains due to their large data requirements, inability to produce physically consistent results, and their lack of generalizability to out-of-sample scenarios \cite{caldwell2014statistical,lazer2014parable,marcus2014eight}. Given that neither an ML-only nor a scientific knowledge-only approach can be considered sufficient for complex scientific and engineering applications, the research community is beginning to explore the continuum between mechanistic and ML models, where both scientific knowledge and ML are integrated in a synergistic manner \cite{karpatne2017theory,karniadakis2021physics,karpatne2022knowledge}. This paradigm is fundamentally different from mainstream practices in the ML community for making use of domain-specific knowledge in feature engineering or post-processing, as it is focused on approaches that integrate scientific knowledge directly into the ML framework.

Even though the idea of integrating scientific principles and ML models has picked up momentum just in the last few years, there is already a vast amount of work on this topic. For instance, in all Web of Science databases (\protect\url{www.webofknowledge.com}) a search for "physics-informed ML" alone shows the growth of publications from 2 in 2017, 8 in 2018, 27 in 2019, to 63 in 2020. Also, many workshops and symposiums have formed to focus on this field (e.g. \protect\cite{iclr2020diff,aaai2020,aaai2021,kgml2020,igars2020,icerm2019,baker2019workshop}). This work is being pursued in diverse disciplines (e.g., earth systems \cite{reichstein2019deep}, climate science \cite{krasnopolsky2006complex,faghmous2014big,o2018using}, turbulence modeling \cite{mohan2018deep,bode2021using,xiao2019reduced}, computational physics \cite{tanaka2021deep}, cyberphysical systems \cite{rai2020driven}, material discovery \cite{raccuglia2016machine,cang2018improving,schleder2019dft}, quantum chemistry \cite{sadowski2016synergies,schutt2017schnet}, biological sciences \cite{yazdani2019systems,peng2021multiscale,alber2019integrating}, and hydrology \cite{xu2015data,varadharajan2021using}), and it is being performed in the context of diverse objectives specific to these applications. Early results in isolated and relatively simple scenarios are promising, and the expectations are rising for this paradigm to accelerate scientific discovery and help address some of the biggest challenges that are facing humanity in terms of climate \cite{faghmous2014big}, health \cite{wang2020tdefsi}, and food security \cite{jia2019bringing}.

The goal of this survey is to bring these exciting developments to the ML community, to make them aware of the progress that has been made, and the gaps and opportunities that exist for advancing research in this promising direction.  We hope that this survey will also be valuable for scientists who are interested in exploring the use of ML to enhance modeling in their respective disciplines. Please note that work on this topic has been referred to by other names, such as "physics-guided ML," "physics-informed ML," or "physics-aware AI" although it covers many scientific disciplines. In this survey, we also use the terms "physics-guided" or "physics," which should be more generally interpreted as "scientific knowledge-guided" or "scientific knowledge".

The focus of this survey is on approaches that integrate scientific knowledge with ML for environmental and engineering systems where scientific knowledge is available as mechanistic models, theories, and laws (e.g. conservation of mass). This distinguishes our survey from other works that focus on more general knowledge integration into machine learning \cite{von2020informed,aditya2019integrating} and works covering physics integration into ML in specific domains (e.g., cyber-physical systems \cite{rai2020driven}, chemistry \cite{noe2020machine}, medical imaging \cite{liu2021anatomy}, fluid mechanics \cite{cai2021physics}, climate and weather\cite{kashinath2021physics}).  This survey creates a novel taxonomy  that covers a wide array of physics-ML methodologies, application-centric objectives, and general computational objectives. 

We organize the paper as follows. Section \protect\ref{sect:objectives} describes different application-centric objectives that are being pursued through combinations of scientific knowledge and ML. Section \protect\ref{sect:methods} discusses novel ML loss functions, model initializations, architectures, and hybrid models that researchers are developing to achieve these objectives, as well as comparisons between them. Section \protect\ref{sec:discussion} discusses the areas of current work as well as the possibilities for cross-fertilization between methods and application-centric objectives. Then, Section \protect\ref{sec:conclusion} contains concluding remarks. Table \protect\ref{table:lit_matrix}  categorizes the work surveyed in this paper by application-centric objective and methods for integrating scientific knowledge in ML according to the proposed taxonomy.

\section{Application-centric Objectives of Physics-ML Integration}
\label{sect:objectives}
\begin{figure}
	\centering
	\includegraphics[width=0.5\linewidth]{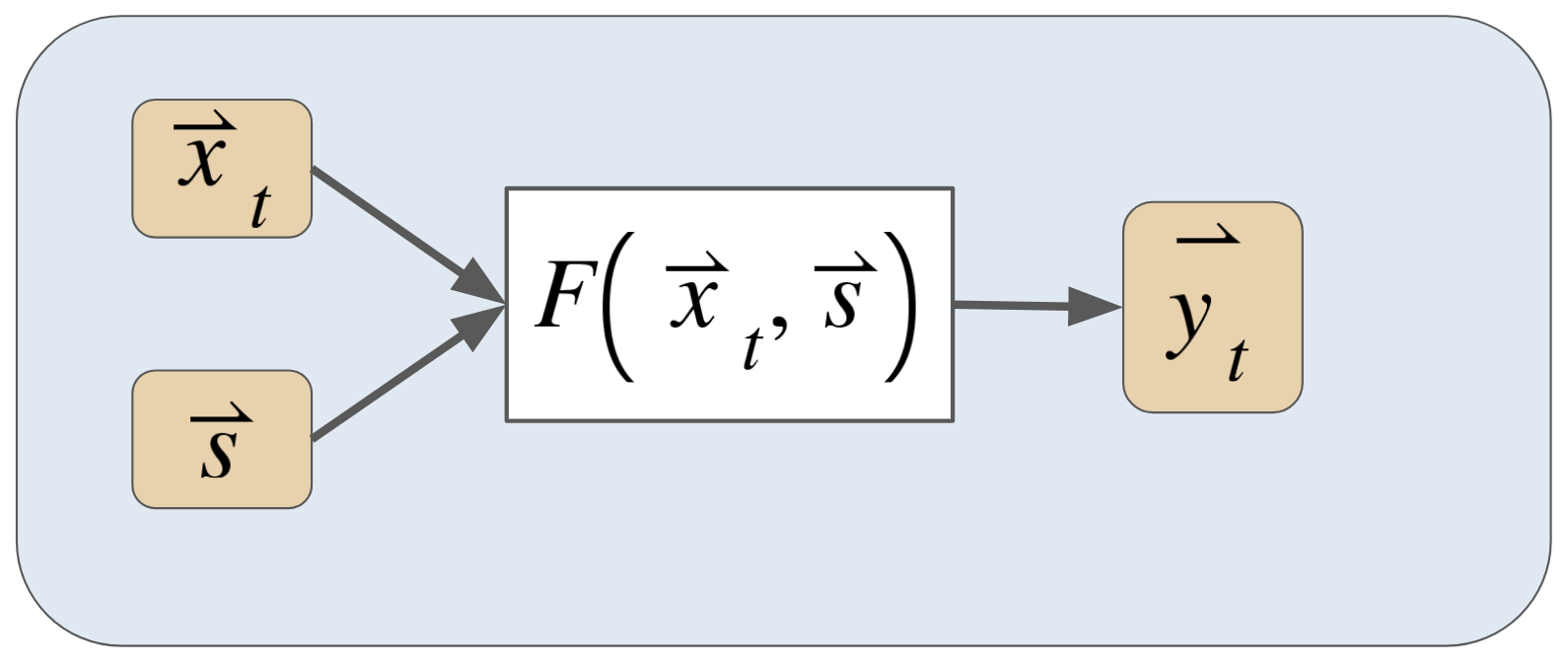}
	\caption{A generic scientific problem in engineering, where $\Vec{x}_t$ are the dynamic inputs in time, $\Vec{s}$ is the set of static characteristics or parameters of the system, and $F()$ is the model producing target variable $\Vec{y}_t$. $\Vec{x}_t$ and $\Vec{y}_t$ can also have spatial dimensions.}
	\label{fig:problem_diagram}
\end{figure}

This section provides a brief overview of a set of application-centric objectives where couplings of ML and scientific modeling paradigms are being pursued for applications in  environmental and engineering systems. In many of these applications, scientific knowledge is represented using a mechanistic model (also referred to as physical, process-based, or first principles models). This is shown in Figure \protect\ref{fig:problem_diagram} as part of an abstract representation of a generic scientific problem. For example, a model for lake temperature would have drivers $\Vec{x}_{t}$, such as amount of sunlight, air temperature, and wind speed, with static parameters $\Vec{s}$, such as lake depth and water clarity, which the model $F(\Vec{x}_{t},\Vec{s})$ would use to predict the water temperature $\Vec{y}_{t}$ at various depths in the lake. Such physical models typically have a notion of state (which is not depicted in Figure 1 for simplicity), and complex physical models can have multiple components that model various aspects of the system, e.g., components to model clouds or ocean in a global climate model. The application-centric objectives described in this section describe different ways in which physics-ML integration can be used to address the imperfections of the mechanistic model $F()$,  build a more resource efficient $F()$, or discover new knowledge such as $F()$. Below we describe how different application-centric objectives fit into each of these scenarios. 


\begin{table}
    \centering
    \scalebox{0.82}{
    \begin{tabular}{|c|c|c|}
        \toprule
         {\rotatebox[origin=c]{\rs}{\large \thead{Objective Name}}} &
         {\rotatebox[origin=c]{\rs}{\large \thead{Objective  Description}}} &
         {\rotatebox[origin=c]{\rs}{\large \thead{Needs}}}\\
         \midrule
         \thead{2.1 Improving Over \\Physical Model}& \thead{Improved version of $F$ that better \\matches observations}& \thead{Observations (synthetic samples \\can be used but not required)} \\
         \midrule
         \thead{2.2 Downscaling\\ \\ \\}&  \thead{An approximate version of $F$ \\that provides high resolution output $Y_{t}$\\ given coarse resolution input $X_{t}$}&  \thead{Synthetic samples at high resolution \\(can also make use of observations at high \\resolutions but
this can be hard to obtain)} \\
         \midrule
         \thead{2.3 Parameterization\\ \\ }&  \thead{Replace a component of $F$ when $F$ consists \\of
interconnected component models} &  \thead{Synthetic samples (e.g. subprocess)\\ \\ } \\
         \midrule
         \thead{2.4 Reduced-\\Order Models\\ \\} &  \thead{Simpler version of $F$ that is\\ more computationally efficient\\ and possibly less accurate}&  \thead{Governing equations (e.g. high-fidelity, \\complex models)\\ \\}\\ 
         \midrule
         \thead{2.5 Forward Solving\\PDEs} &  \thead{Computationally efficient single \\pass solution of PDE }&  \thead{Governing equations (e.g. high-fidelity, \\complex models)}\\
         \midrule
         \thead{2.6 Inverse Modeling\\ \\}&  \thead{Discover static characteristics $S$ given \\$\Vec{x}_t$ and $\Vec{y}_t$}&  \thead{Observations (synthetic samples \\can be used but not required)} \\
         \midrule
         \thead{2.7 Discovering \\Governing Equations}&  \thead{Find governing equations that underlie $F$\\ \\}&  \thead{Observations\\ \\} \\
         \midrule
         \thead{2.8 Data Generation\\ \\}& \thead{Generate realistic synthetic samples\\ without using $F$}&  \thead{Synthetic data (or observations)\\ \\}   \\
         \midrule
         \thead{2.9 Uncertainty \\Quantification\\} &  \thead{Estimate uncertainty on $\Vec{y}_t$ given input $\Vec{x}_t$\\ \\}&  \thead{Observations or synthetic samples\\ \\}\\
         \bottomrule
         \bottomrule
    \end{tabular}}
	\caption{Application-centric objectives of using Physics-ML methods in terms of the generic scientific problem shown in Figure \protect\ref{fig:problem_diagram} and the needs to pursue each objective}
	\label{tab:objective_directions}
\end{table}

First, situations can arise in which a mechanistic model is inadequate to model a poorly understood process and a data-driven model could be leveraged to make better use of observations and possibly also process-based knowledge. Section \protect\ref{obj:improve_over_phys} covers approaches like this, where physics-ML is being pursued to improve the effectiveness and predictive accuracy of $F(\Vec{x}_{t},\Vec{s})$ with observations and scientific knowledge. 

Often it is also desirable to improve the resource efficiency where physical models are too slow or at too coarse of a resolution. Section \protect\ref{obj:downscaling} on downscaling considers how physics-ML can produce high-resolution output variables faster than a physical model in situations spanning meteorology, climatology, and others. Similarly, if subprocesses of a larger mechanistic model are computationally intractable or inaccurate, an ML model can be used for faster or more accurate subprocess representation, as is covered in Section \protect\ref{obj:parameterization} on parameterization. More generally, the concept of reducing computational complexity of complex mechanistic or numerical models in the form of an ML-based or ML-assisted reduced order model is described in Section \protect\ref{obj:roms}. Another objective seeking to improve resource efficiency with physics-ML is the solving of partial differential equations (PDEs) where the solution is represented with a data-driven model. This allows dynamical systems applications to bypass the often extreme computational complexity of using finite elements methods to solve complex systems of PDEs. As we can see, in many of these cases ML can be used as a more efficient \textit{surrogate model} (also referred to as emulators) for an existing mechanistic or numerical approach. 

Other objectives seek to discover new knowledge in the form of unobserved causal quantities or the symbolic representation of a process given only data. Compared to previous objectives where the goal is to produce accurate or efficient output variables, inverse modeling described in Section \protect\ref{obj:inverse} flips the path shown in Figure \protect\ref{fig:problem_diagram} and instead seeks to find some causal static physical parameters within $\Vec{s}$ given sufficient outputs. Also, Section \protect\ref{obj:govern_eq} covers the discovering of governing equations from data, where ML has been used to go beyond traditional approaches and discover new explicit symbolic representations of phenomena.

Data generation objectives (Section \protect\ref{obj:data_gen}) try to realistically reproduce the distribution of $\Vec{x}_t$, $\Vec{y}_t$, or ($\Vec{x}_t$, $\Vec{y}_t$). Such synthetically generated data can be useful in the often data-limited situations present in engineering and environmental systems. UQ (Section \ref{obj:uq}) tries to learn the distribution of $\Vec{y}_{t}$ in terms of the uncertainty of the other components of the model like the inputs, static parameters, and model state. UQ is necessary for accurate forecasting, which can also affect many of the other objectives.  

Table \protect\ref{tab:objective_directions} summarizes these objectives and their needs in terms of real world observations, synthetic samples (e.g. output from a mechanistic model), or knowledge of the governing equations of the system.

\subsection{Improving over state-of-the-art physical models}
\label{obj:improve_over_phys}

First-principle models are used extensively in a wide range of engineering and environmental applications. Even though these models are based on known physical laws, in most cases, they are necessarily approximations of reality due to incomplete knowledge of certain processes, which introduces bias. In addition, they often contain a large number of parameters whose values must be estimated with the help of limited observed data, degrading their performance further, especially due to heterogeneity in the underlying processes in both space and time. The limitations of physics-based models cut across discipline boundaries and are well known in the scientific community (e.g., see Gupta et al.  \cite{gupta2014debates} in the context of hydrology).

ML models have been shown to outperform physics-based models in many disciplines (e.g., materials science \cite{kauwe2018machine,wei2018predicting,ryan2018crystal}, applied physics \cite{ibarra2015short,baldi2016jet}, aquatic sciences \protect\cite{jia2020physics,willard2021predicting}, atmospheric science \cite{nowack2018using}, biomedical science \cite{tesche2018coronary}, computational biology \cite{alipanahi2015predicting}). A major reason for this success is that ML models (e.g., neural networks), given enough data, can find structure and patterns in problems where complexity prohibits the explicit programming of a system's exact physical nature. Given this ability to automatically extract complex relationships from data, ML models appear  promising for scientific problems with physical processes that are not fully understood but have data of adequate quality and quantity is available. However, the black-box application of ML has met with limited success in scientific domains due to a number of reasons \cite{karpatne2017theory}: (i) while state-of-the-art ML models are capable of capturing complex spatiotemporal relationships, they require far too much labeled data for training, which is rarely available in real application settings, (ii) ML models often produce scientifically inconsistent results; and (iii) ML models can only capture relationships in the available training data, and thus cannot generalize to out-of-sample scenarios (i.e., those not represented in the training data).

The key objective here is to combine elements of physics-based modeling with state-of-the-art ML models to leverage their complementary strengths. Such integrated physics-ML models are expected to better capture the dynamics of scientific systems and advance the understanding of underlying physical processes. An early attempt for combining ML with physics-based modeling in lake  temperature  dynamics  \protect\cite{read2019process}) has  already  demonstrated  its  potential  for  providing  better  prediction  accuracy  with  a much  smaller  number  of  samples  as  well as  generalizability  in  out-of-sample   scenarios.


\subsection{Downscaling}
\label{obj:downscaling}
Complex mechanistic models are capable of capturing physical reality more precisely than simpler models, as they often involve more diverse components that account for a greater number of processes at finer spatial or temporal resolution. However, given the computation cost and modeling complexity, many models are run at a resolution that is coarser than what is required to precisely capture underlying physical processes. For example, cloud-resolving models (CRM) need to run at sub-kilometer horizontal resolution to be able to effectively represent boundary-layer eddies and low clouds, which are crucial for the modeling of Earth’s energy balance and the cloud–radiation feedback~\cite{rasp2018deep}. However, it is not feasible  to run global climate models at such fine resolutions even with the most powerful computers expected to be available in the near future.

Downscaling techniques have been widely used as a solution to capture physical variables that need to be modeled at a finer resolution. In general, the downscaling techniques can be divided into two categories: statistical downscaling and dynamical downscaling. 
Statistical downscaling refers to the use of empirical models to predict finer-resolution variables from coarser-resolution variables. Such a mapping across different resolutions can involve complex non-linear relationships that cannot be precisely modeled by traditional empirical models.  Recently, artificial neural networks have shown a lot of promise  for this problem, given their ability to model non-linear relationships~\cite{vandal2017deepsd,sharifi2019downscaling}. In contrast, dynamical downscaling  makes use of high-resolution regional simulations to dynamically simulate relevant physical processes at regional or local scales of interest. Due to the substantial time cost of running such complex models, there is an increasing interest in using ML models as surrogate models (a model approximating simulation-driven input–output data) to predict target variables at a higher resolution~\cite{srivastava2013machine,gentine2018could}. 

Although the state-of-the-art ML methods can be used in both statistical and dynamical downscaling, it remains a challenge to ensure that the learned ML component is consistent with established physical laws and can improve the overall simulation performance.

\subsection{Parameterization}
\label{obj:parameterization}

Complex physics-based models (e.g., for simulating phenomena in climate, weather, turbulence modeling, hydrology) often use an approach known as \textit{parameterization} to account for missing physics. In parameterization (note that this term has a different meaning when used in mathematics and geometry), specific complex dynamical processes are replaced by simplified physical approximations that are represented as static parameters. A common way to estimate the value of these parameters is to use grid search over the space of combinations of parameter values that lead to the best match with observations. This procedure is referred to as parameter calibration. The failure to correctly parameterize can make the model less robust, and errors that result from imperfect parameterization can also feed into other components of the entire physics-based model and deteriorate the modeling of important physical processes. Another way, that is being considered increasingly, is to replace processes that are too complex to be physically represented in the model by a simplified dynamic or statistical/ML process.  This makes it possible to learn new parameterizations directly from observations and/or high-resolution model simulation using ML methods. Already, ML-based parameterizations have shown success in geology \cite{goldstein2014data,chan2017parametrization}, atmospheric science \cite{gentine2018could,brenowitz2018prognostic}, and hydrology \cite{bennett2020deep}. A major benefit of ML-based parameterizations is the reduction of computation time compared to traditional physics-based simulations. In chemistry, Hansen et al. \cite{hansen2013assessment} find that parameterizations of atomic energy using ML take seconds compared to multiple days for more standard quantum-calculation calculations, and Behler et al. \cite{behler2011neural} find that neural networks can vastly improve the efficiency of finding potential energy surfaces of molecules.

Most of the existing work uses standard black box ML models for parameterization, but there is an increasing interest in integrating physics in the ML models \cite{beucler2019enforcing}, as it has the potential to make them more robust and generalizable to unseen scenarios as well as reduce the number of training samples needed for training.

\subsection{Reduced-Order Models}
\label{obj:roms}

Reduced-Order Models (ROMs) are computationally inexpensive representations of more complex models. Usually, constructing ROMs involves dimensionality reduction that attempts to capture the most important dynamical characteristics of often large, high-fidelity simulations and models of physical systems (e.g., in fluid dynamics \cite{lassila2014model}). This can also be viewed as a controlled loss of accuracy. A common way to do this is to project the governing equations of a system onto a linear subspace of the original state space using a method such as principal components analysis or dynamic mode decomposition \cite{quarteroni2014reduced}. However, limiting the dynamics to a lower-dimensional subspace inherently limits the accuracy of any ROM.  

ML is beginning to assist in constructing ROMs for increased accuracy and reduced computational cost in several ways. One approach is to build an  ML-based surrogate model for full-order models \cite{chen2012support,kasim2020building}, where the ML model can be considered a ROM. Other ways include building an ML-based surrogate model of an already built ROM by another dimensionality reduction method \cite{xiao2019reduced} or building an ML model to mimic the dimensionality reduction mapping from a  full-order model to a reduced-order model \cite{mohan2018deep}. ML and ROMs can also be combined by using the ML model to learn the residual between a ROM and observational data \cite{wan2018data}. ML models have the potential to greatly augment the capabilities of ROMs because of their typically quick forward execution speed and ability to leverage data to model high dimensional phenomena. 

One area of recent focus of ML-based ROMs is in approximating the dominant modes of the Koopman (or composition) operator, as a method of dimensionality reduction. The Koopman operator is an infinite-dimension linear operator that encodes the temporal evolution of the system state through nonlinear dynamics \cite{brunton2017koopman}. This allows linear analysis methods to be applied to nonlinear systems and enables the inference of properties of dynamical systems that are too complex to express using traditional analysis techniques. Applications span many disciplines, including fluid dynamics \cite{sharma2016correspondence}, oceanography \cite{giannakis2015spatiotemporal}, molecular dynamics \cite{wu2017variational}, and many others. Though dynamic mode decomposition \cite{mezic2005spectral} is the most common technique for approximating the Koopman operator, many recent approaches have been made to approximate Koopman operator embeddings with deep learning models that outperform existing methods \cite{li2017extended,takeishi2017learning,lusch2018deep,wehmeyer2018time,mardt2018vampnets,morton2019deep,otto2019linearly,yeung2019learning,pan2020physics}. Adding physics-based knowledge to the learning of the Koopman operator has the potential to augment generalizability and interpretability, which current ML methods in this area tend to lack  \cite{pan2020physics}.

Traditional methods for ROMs often lack robustness with respect to parameter changes within the systems they are representing \cite{amsallem2008interpolation}, or are not cost-effective enough when trying to predict complex dynamical systems (e.g., multiscale in space and
time). Thus, incorporating principles from physics-based models could potentially reduce the search space to enable more robust training of ROMs, and also allow the model to be trained with less data in many scenarios. 

\subsection{Forward Solving Partial Differential Equations}
\label{obj:solve_pde}
In many physical systems, governing equations are known, but direct numerical solutions of partial differential equations (PDEs) using common methods, such as the Finite Elements Method or the Finite Difference Method \cite{jacob2007first}, are  prohibitively expensive. In such cases, traditional methods are not ideal or sometimes even possible. A common technique is to use an ML model as a surrogate for the solution to reduce computation time \cite{dissanayake1994neural,lagaris1998artificial}. In particular, NN solvers can reduce the high computational demands of traditional numerical methods into a single forward-pass of a NN. Notably,  solutions obtained via NNs are also naturally differentiable and have a closed analytic form that can be transferred to any subsequent calculations, a feature not found in more traditional solving methods \cite{lagaris1998artificial}. Especially with the recent advancement of computational power, neural network models have shown success in approximating solutions across different kinds of physics-based PDEs \cite{arsenault2014machine,rudd2015constrained,khoo2019solving}, including the difficult quantum many-body problem \cite{carleo2017solving} and many-electron Schrodinger equation \cite{han2019solving}. As a step further, deep neural networks models have shown success in approximating solutions across  high dimensional physics-based PDEs previously considered unsuitable for approximation by ML \cite{han2018solving,sirignano2018dgm}. However, slow convergence in training, limited applicability to many complex systems, and reduced accuracy due to unawareness of physical laws can prove problematic. More recently, Li et al. \protect\cite{li2020fourier} have defined a neural Fourier operator which allows a neural network to learn and solve an entire family of PDEs by learning the mapping from any functional parametric dependence to the solution in Fourier space. 

\subsection{Inverse Modeling}
\label{obj:inverse}
The forward modeling of a physical system uses the physical parameters of the system (e.g., mass, temperature, charge, physical dimensions or structure) to predict the next state of the system or its effects (outputs). In contrast,  inverse modeling uses the (possibly noisy) output of a system to infer the intrinsic physical parameters or inputs. Inverse problems often stand out as important in physics-based modeling communities because they can potentially shed light on valuable information that cannot be observed directly. One example is the use of x-ray images from a CT scan to create a 3D image reflecting the structure of part of a person's body \cite{lunz2018adversarial}. This can be viewed as a computer vision problem, where, given many training datasets of x-ray scans of the body at different capture angles, a model can be trained to inversely reconstruct textures and 3D shapes of different organs or other areas of interest. Allowing for better reconstruction while reducing scan time could potentially increase patient satisfaction and reduce overall medical costs. Though there are many inverse modeling scenarios, in this work we focus on intrinsic physical parameters found in a mechanistic modeling scenario for engineering and environmental systems. 

Often, the solution of an inverse problem can be computationally expensive due to the potentially millions of forward model evaluations needed for estimator evaluation or characterization of posterior distributions of physical parameters \cite{frangos2010surrogate}. ML-based surrogate models (in addition to other methods such as reduced-order models) are becoming a realistic choice since they can model high-dimensional phenomena with lots of data and execute much faster than most physical simulations.

Inverse problems are traditionally solved using regularized regression techniques.  
Data-driven methods have seen success in inverse problems in remote sensing of surface properties \cite{dawson1992inversion}, hydrology \protect\cite{ghorbanidehno2020recent}, photonics \cite{pilozzi2018machine}, and medical imaging \cite{lunz2018adversarial}, among many others.
Recently, novel algorithms using deep learning and neural networks have been applied to inverse problems. While still in their infancy, these techniques exhibit strong performance for applications such as computerized tomography \cite{chen2017low,mccann2017review,tayal2020unlocking}, seismic processing \cite{vamaraju2019unsupervised}, or various sparse data problems. 

There is also increasing interest in the inverse design of materials using ML, where desired target properties of materials are used as input to the model to identify atomic or microscale structures that exhibit such properties \cite{raccuglia2016machine,schleder2019dft,liao2020metaheuristic,kumar2020inverse}. Physics-based constraints and stopping conditions based on material properties can be used to guide the optimization process \cite{liao2020metaheuristic,tayal2020inverse}. These constraints and similar physics-guided techniques have the potential to alleviate noted challenges in inverse modeling,  particularly in scenarios with a small sample size and a paucity of ground-truth labels  \cite{karpatne2018machine}. 
The integration of prior physical knowledge is common in current approaches to the inverse problem, and its integration into ML-based inverse models has the potential to improve data efficiency and increase its ability to solve ill-posed inverse problems.

\subsection{Discovering Governing Equations}
\label{obj:govern_eq}

When the governing equations of a dynamical system are known explicitly, they allow for more robust forecasting, control, and the opportunity for analysis of system stability and bifurcations through increased interpretability \cite{rudy2019deep}. Furthermore, if a mathematical model accurately describes the processes governing the observed data, it therefore can generalize to data outside of the training domain.  However, in many disciplines (e.g., neuroscience, cell biology, ecology,  epidemiology)  dynamical systems have no  formal analytic descriptions. Often in these cases, data is abundant, but the underlying governing equations remain elusive. In this section, we discuss equation discovery systems that do not assume the structure of the desired equation (as in Section \ref{obj:solve_pde}), but rather explore a large space of possibly nonlinear mathematical terms.

Advances in ML for the discovery of these governing equations has become an active research area with rich potential to integrate principles from applied mathematics and physics with modern ML methods. Early works on the data-driven discovery of physical laws relied on heuristics and expert guidance and were focused on rediscovering known, non-differential, laws in different scientific disciplines from artificial data \cite{gerwin1974information,langley1981data,langley1983rediscovering,lenat1983role}. This was later expanded to include real-world data and differential equations in ecological applications \cite{dvzeroski1999equation}. Recently, general and robust data-driven discovery of potentially unknown governing equations has been pioneered by \cite{bongard2007automated,schmidt2009distilling}, where they apply symbolic regression to differences between computed derivatives and analytic derivatives to determine underlying dynamical systems. More recently, works have used sparse regression built on a dictionary of functions and partial derivatives to construct governing equations \cite{brunton2016discovering,rudy2017data,quade2018sparse}. Lagergren et al.  \cite{lagergren2020learning} expand on this by using ANNs to construct the dictionary of functions. These sparse identification techniques are based on the principle of Occam's Razor, where the goal is that only a few equation terms be used to describe any given nonlinear system.

\subsection{Data Generation}
\label{obj:data_gen}
Data generation approaches are useful for creating virtual simulations of scientific data under specific conditions. For example, these techniques can be used to generate potential chemical compounds with desired characteristics (e.g., serving as catalysts or having a specific crystal structure). Traditional physics-based approaches for generating data often rely on  running physical simulations or conducting physical experiments, which tend to be very  time consuming. Also, these approaches are restricted by what can be produced by physics-based models. Hence, there is an increasing interest in generative ML approaches that learn data distributions in unsupervised settings and thus have the potential to generate novel data beyond what could be produced by traditional approaches. 

Generative ML models have found tremendous success in areas such as speech recognition and generation \cite{oord2016wavenet}, image generation \cite{denton2015deep}, and natural language processing \cite{guo2018long}.  
These models have been at the forefront of unsupervised learning in recent years, mostly due  to their efficiency in understanding unlabeled data. The idea behind generative models is to capture the inner probabilistic distribution 
in order to generate similar data. 
With the recent advances in deep learning, new generative models, such as the generative adversarial network (GAN) and variational autoencoder (VAE), have been developed.  These models have shown much better performance in learning non-linear relationships to extract representative latent embeddings from observation data. Hence the data generated from the latent embeddings are more similar to true data distribution. 
In particular, the adversarial component of GAN consists of a two-part framework with a generative network and discriminative network, where the generative network's objective is to generate fake data to fool the discriminative network, while the discriminative network attempts to determine true data from fake data.

In the scientific domain, GANs can generate data like the data  generated by the physics-based models. Using GANs often incurs certain benefits, including reduced computation time and  a better reproduction of complex phenomenon, given the ability of GANs to represent nonlinear relationships. 
For example, Farimani et al. ~\cite{farimani2017deep} have shown that Conditional Generative Adversarial Networks (cGAN) can be trained to simulate heat conduction and fluid flow purely based on observations without using knowledge of the underlying governing equations. Such approaches that use generative models have been shown to significantly accelerate the process of generating new data samples.

However, a well-known issue of GANs is that they incur dramatically high sample complexity. Therefore, a growing area of research is to engineer GANs that can leverage prior knowledge of physics in terms of physical laws and invariance properties. 
For example, GAN-based models for simulating turbulent flows can be further improved  by incorporating physical constraints, e.g., conservation laws~\cite{zeng2019enforcing} and  energy spectrum~\cite{wu2019enforcing}, in the loss function. Cang et al.~\cite{cang2018improving} also imposed a physics-based morphology constraint on a VAE-based generative model used for simulating artificial material samples.  The physics-based constraint forces the generated artificial samples to have the same morphology distribution as the authentic ones and thus greatly reduces the large material design space.

\subsection{Uncertainty Quantification}
\label{obj:uq}
Uncertainty quantification (UQ) is of great importance in many areas of computational science (e.g., climate modeling \cite{deser2012uncertainty}, fluid flow \cite{christie2006uncertainty}, systems engineering \cite{pettit2004uncertainty}, among many others). At its core, UQ requires an accurate characterization of the entire distribution $p(y|x)$, where $y$ is the response and $x$ is the covariate of interest, rather than just making a point prediction $y = f(x)$. This makes it possible to characterize all quantiles and skews in the distribution, which allows for analysis such as examining how close predictions are to being unacceptable, or sensitivity analysis of input features.

Applying UQ tasks to physics-based models using traditional methods such as Monte Carlo (MC) is usually infeasible due to the very large number of forward model evaluations needed to obtain convergent statistics. In the physics-based modeling community, a common technique is to perform model reduction (described in Section \ref{obj:roms}) or create an ML surrogate model, in order to increase model evaluation speed since ML models often execute much faster \cite{galbally2010non,manzoni2016accurate,tripathy2018deep}. With a similar goal, the ML community has often employed Gaussian Processes as the main technique for quantifying uncertainty in simulating physical processes \cite{bilionis2012multi,rajabi2017uncertainty}, but neither Gaussian Processes nor reduced models scale well to higher dimensions or larger datasets (Gaussian Processes scale as $\mathscr O(N^3)$ with $N$ data points). 

Consequently, there is an effort to fit deep learning models, which have exhibited countless successes across disciplines, as a surrogate for numerical simulations in order to achieve faster model evaluations for UQ that have greater scalability than Gaussian Processes \cite{tripathy2018deep}. However, since artificial neural networks do not have UQ naturally built into them, variations have been developed. One such modification uses a probabilistic drop-out strategy in which neurons are periodically "turned off"  as a type of Bayesian approximation to estimate uncertainty ~\cite{gal2016dropout}. There are also Bayesian variants of neural networks that consist of distributions of weights and biases \cite{mackay1992practical,zhang2009estimating,zhu2018bayesian}, but these suffer from high computation times and high dependence on reliable priors. Another method uses an ensemble of neural networks to create a distribution from which uncertainty is quantified \cite{lakshminarayanan2017simple}.


The integration of physics knowledge into ML for UQ has the potential to allow for a better characterization of uncertainty. For example, ML surrogate models run the risk of producing physically inconsistent predictions, and incorporating elements of physics could help with this issue. Also, note that the reduced data needs of ML due to constraints for adherence to known physical laws could  alleviate some of the high computational cost of Bayesian neural networks for UQ. 




\section{Physics-ML Methods}
\label{sect:methods}
Given the diversity of forms in which scientific knowledge is represented in different disciplines and applications, researchers have developed a large variety of methods for integrating physical principles into ML models. This section categorizes them into the following four classes; (i) physics-guided loss function, (ii) physics-guided initialization, (iii) physics-guided design of architecture, and (iv) hybrid modeling.

Choosing between different classes of methods for a given problem can depend on many factors including the availability and performance of existing mechanistic models, and also the general computational objectives that need to be addressed. The general computational objectives of physics-ML methods described throughout this section, as opposed to traditional ML methods, can be placed into three categories. First, \textit{prediction performance} defined as better matching between predicted and observed values can be improved in a variety of ways including improved generalizability to out-of-sample scenarios,  improved general accuracy, or forcing solutions to be physically consistent (e.g. obeying known physics-based governing equations). Second, \textit{sample efficiency} can be improved by reducing the number of observations required for adequate performance or reducing the overall search space. The third general computational objective is \textit{interpretability}, where often traditional ML models are a "black box" and the incorporation of scientific knowledge can shine light on physical meanings, interpretations, and processes within the ML framework. Even though computational objectives can be categorized within these categories, there is also overlap between them. For example, forcing models to be physically consistent can effectively reduce the solution search space. Improved sample efficiency can also lead to improved prediction performance by getting more value out of each observation. We end this section with a summary and detailed discussion comparing different kinds of methods, their requirements, and the general computational objectives achieved.

\subsection{Physics-Guided Loss Function}
\label{method:loss_func}

Scientific problems often exhibit a high degree of complexity due to relationships between many physical variables varying across space and time at different scales. Standard ML models can fail to capture such relationships directly from data, especially when provided with limited observation data. This is one reason for their failure to generalize to scenarios not encountered in training data. Researchers are beginning to incorporate physical knowledge into loss functions to help ML models capture generalizable dynamic patterns consistent with established physical laws. 

One of the most common techniques to make ML models consistent with physical laws is to incorporate physical constraints into the loss function of ML models as follows \cite{karpatne2017theory},
\begin{equation}\label{eq1}
    \text{Loss} = \text{Loss}_{\text{TRN}}(Y_{\text{true}}, Y_{\text{pred}}) + \lambda R(W) + \gamma\text{Loss}_{\text{PHY}}(Y_{\text{pred}})
\end{equation} where the training loss $\text{Loss}_{\text{TRN}}$ measures a supervised error (e.g., RMSE or cross-entropy) between true labels $Y_{\text{true}}$ and predicted labels $Y_{\text{pred}}$, and $\lambda$ is a hyper-parameter to control the weight of model complexity loss $R(W)$. The first two terms are the standard loss of ML models.  The addition of physics-based loss $\text{Loss}_{\text{PHY}}$ aims to ensure consistency with physical laws and it is weighted by a hyper-parameter $\gamma$, where $\gamma$ is determined alongside other ML hyperparameters using validation data or a nested cross validation setup. A comprehensive guide to implementing physics-based loss functions can be found in Ebert-Uphoff et al. \cite{ebert2021cira}.

Steering ML predictions towards physically consistent outputs has numerous benefits. First, this provides the possibility to ensure the consistency with physical laws and therefore reduce the solution search space of ML models. Second, the regularization by physical constraints allows the model to learn even with unlabeled data, as the computation of  physics-based loss \protect{$\text{Loss}_{\text{PHY}}$}  does not require observation data. Third, ML models which  follow desired physical properties are more likely to be generalizable to out-of-sample scenarios relative to basic ML models\protect\cite{jia2021physics,read2019process}. It is important to note, however, that the physics-guided loss function does not "guarantee" either physical consistency or generalizability as it is fundamentally a weak constraint. Loss function terms corresponding to physical constraints are applicable across many different types of ML frameworks. In addition, this method is extensively used across all application-centric objectives listed in Section \protect\ref{sect:objectives}. In the following paragraphs, we demonstrate the use of physics-based loss functions for different objectives described in Section \protect\ref{sect:objectives}.

\paragraph{Replacing or improving over physical models}
Incorporation of physics-based loss has shown great success in improving prediction ability of ML models. In the context of lake temperature modeling, Karpatne et al. \cite{karpatne2017physics} includes a physics-based penalty that ensures that predictions of denser water are at lower depths than predictions of less dense water, a known monotonic relationship.

\begin{figure}
    \centering
    \includegraphics[width=1\linewidth]{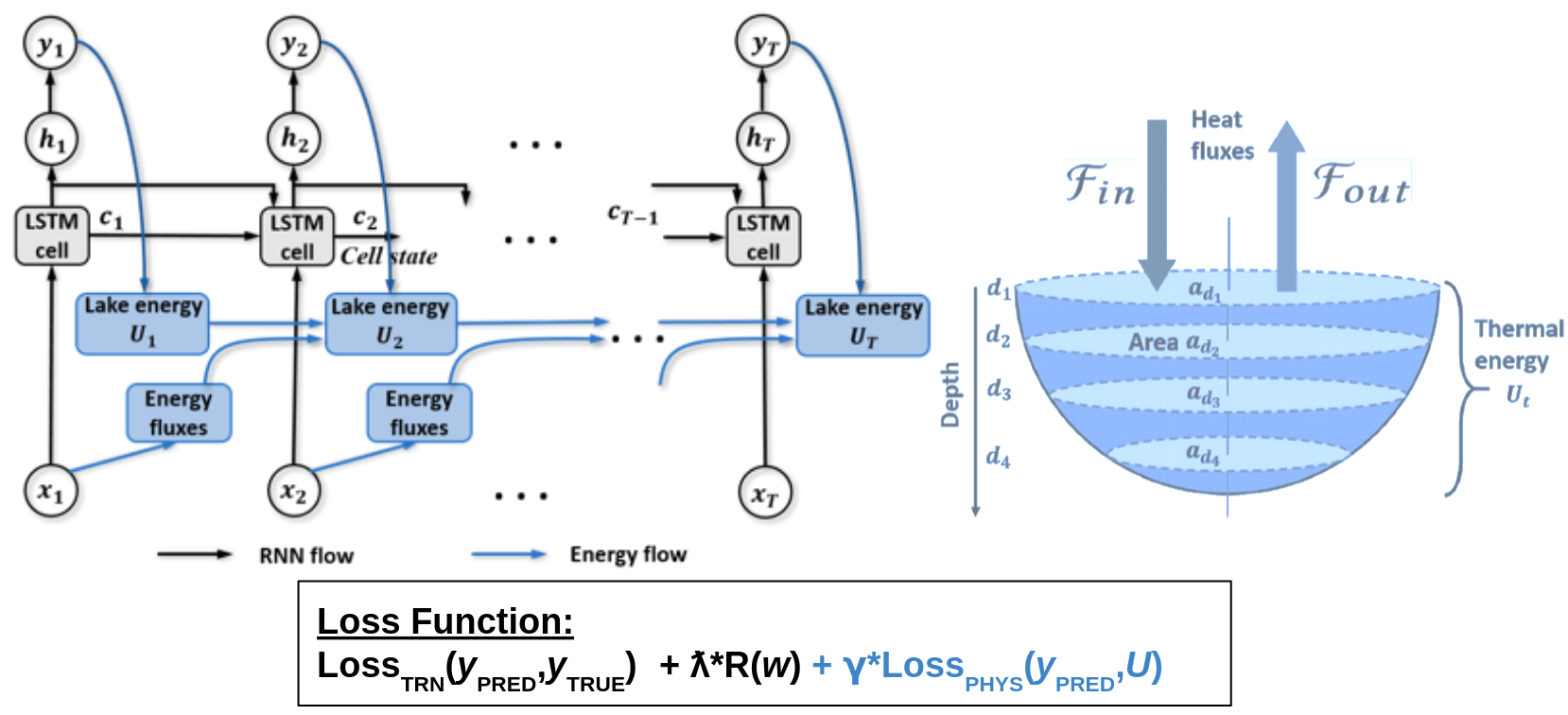}
    \caption{The Physics-Guided Recurrent Neural Network (PGRNN) model demonstrated in Jia et al. \protect\cite{jia2020physics} is an example of a physics-guided loss function allowing physical knowledge to be incorporated into the ML model. They include the standard RNN flow (black arrows) and the energy flow (blue arrows) in the recurrent process. Here $U_T$ represents the thermal energy of the lake at time $T$, and both the energy output and temperature output $y_T$ are used in calculating the loss function value. This enables the PGRNN to predict lake temperature without violating energy constraints. A detailed description of the loss function equation (Equation \protect\ref{eq1}) can be found in Section \protect\ref{method:loss_func}. 
}
    \label{fig:pgrnn}
\end{figure}

Jia et al. \cite{jia2019physics} and Read et al. \cite{read2019process} further extended this work to capture even more complex and general physical relationships that happen on a temporal scale. Specifically, they use a physics-based penalty for energy conservation in the loss function to ensure the lake thermal energy gain across time is consistent with the net thermodynamic fluxes in and out of the lake. A diagram of this model is shown in Figure \ref{fig:pgrnn}. Note that the recurrent structure contains additional nodes (shown in blue) to represent physical variables (lake energy, etc) that are computed using purely physics-based equations. These are needed to incorporate energy conservation in the loss function. Similar structure can be used to model other physical laws such as mass conservation, etc. Qualitative mathematical properties of dynamical systems modeling have also shown promise in informing loss functions to improve prediction beyond that of the physics model. Erichson et al. \protect\cite{erichson2019physics}  penalize autoencoders based on physically meaningful stability measures in dynamical systems to improve prediction of fluid flow and sea surface temperature. They showed an improved mapping of past states to future states for both modeling scenarios in addition to improving generalizability to new data.

\paragraph{Solving PDEs}
Another strand of work that involves loss function alterations is solving PDEs for dynamical systems modeling, in which adherence to the governing equations is enforced in the loss function. In Raissi et al. \cite{raissi2019physics}, this concept is developed and shown to create data-efficient spatiotemporal function approximators to both solve and find parameters of basic PDEs like Burgers Equation or Schrodinger Equation. Going beyond a simple feed-forward network, Zhu et al. \cite{zhu2019physics} propose an encoder-decoder model for predicting transient PDEs with governing PDE constraints. Geneva et al. \cite{geneva2020modeling} extended this approach to deep auto-regressive dense encoder-decoders with a Bayesian framework using stochastic weight averaging to quantify uncertainty.

\paragraph{Discovering Governing Equations}

Physics-based loss function terms have also been used in the discovery of governing equations. Loiseau et al. ~\cite{loiseau2018constrained} used constrained least squares \cite{golub2012matrix} to incorporate energy-preserving nonlinearities or to enforce symmetries in the identified equations for the equation learning process described in Section \ref{obj:govern_eq}. Though these loss functions are mostly seen in common variants of NNs, they are also be seen in architectures such as echo state networks. Doan et al. \cite{doan2019physics} found that integrating the physics-based loss from the governing equations in a Lorenz system, a commonly studied system in dynamical systems, strongly improves the echo state network's time-accurate prediction of the system and also reduces convergence time.

\paragraph{Inverse modeling}
For applications in vortex induced vibrations, Raissi et al. \cite{raissi2019deep} pose the inverse modeling problem of predicting the lift and drag forces of a system given sparse data about its velocity field. Kahana et al. \cite{kahana2020obstacle} uses a loss function term pertaining to the physical consistency of the time evolution of waves for the inverse problem of identifying the location of an  underwater obstacle from acoustic measurements. In both cases, the addition of physics-based loss terms made results more accurate and more robust to out-of-sample scenarios.

\paragraph{Parameterization}
While ML has been used for parameterization, adding physics-based loss terms can further benefit this process by ensuring physically consistent outputs.  
Zhang et al. \cite{zhang2018deep} parameterize atomic energy for molecular dynamics using a NN with a loss function that takes into account atomic force, atomic energy, and terms relating kinetic and potential energy. Furthermore, in climate modeling, Beucler \textit{et al.} show
 that enforcing energy conservation laws improves prediction when emulating cloud processes    ~\cite{beucler2019achieving,beucler2019enforcing}.  

\paragraph{Downscaling}
Super resolution and downscaling frameworks have also begun to incorporate physics-based constraints. Jiang et al. \protect\cite{jiang2020meshfreeflownet} use PDE-based constraints for super resolution problems in computational fluid dynamics where they are able to more efficiently recover physical quantities of interest. Bode et al. \protect\cite{bode2021using} use similar constraint ideas in building generative adversarial networks for super resolution in turbulence modeling in combustion scenarios, where they find improved generalization capability and extrapolation due to the constraints.

\paragraph{Uncertainty quantification}
In Yang et al. \cite{yang2018physics2} and Yang et al. \cite{yang2019adversarial}, physics-based loss is implemented in a deep probabilistic generative model for uncertainty quantification based on adherence to the structure imposed by PDEs. To accomplish this, they construct probabilistic representations of the system states, and use an adversarial inference procedure to train using a physics-based loss function that enforces adherence to the governing laws. This is expanded in Zhu et al. \cite{zhu2019physics}, where a physics-informed encoder-decoder network is defined in conjunction with a conditional flow-based generative model for similar purposes. A similar loss function modification is performed in other works \cite{yang2019highly,karumuri2020simulator,geneva2020modeling}, but for the purpose of solving high dimensional stochastic PDEs with uncertainty propagation. In these cases, physics-guided constraints provide effective regularization for training deep generative models to serve as surrogates of physical systems where the cost of acquiring data is high and the data sets are small \cite{yang2019adversarial}.

Another direction for encoding physics knowledge into ML UQ applications is to create a physics-guided Bayesian NN. This is explored in Yang et al. \cite{yang2020b}, where they use a Bayesian NN, which naturally encodes uncertainty, as a surrogate for a PDE solution. Additionally, they add a PDE constraint for the governing laws of the system to serve as a prior for the Bayesian net, allowing for more accurate predictions in situations with significant noise due to the physics-based regularization.

\paragraph{Generative models}
In recent years, GANs have been used to efficiently generate solutions to PDEs and there is interest in using physics knowledge to improve them. Yang et al. \cite{yang2018physics} showed GANs with loss functions based on PDEs can be used to solve stochastic elliptic PDEs in up to 30 dimensions. In a similar vein, Wu et al. \cite{wu2020enforcing} showed that physics-based loss functions in GANs can lower the amount of data and training time needed to converge on solutions of turbulence PDEs, while Shah et al. \cite{shah2019encoding} saw similar results in the generation of microstructures satisfying certain physical properties in computational materials science.

\subsection{Physics-Guided Initialization}
\label{method:init}
Since many ML models require an initial choice of model parameters before training, researchers have explored different ways to physically inform a model starting state. For example, in NNs, weights are often initialized according to a random distribution prior to training. Poor initialization can cause models to anchor in local minima, which is especially true for deep neural networks. However, if physical or other contextual knowledge is used to help inform the initialization of the weights, model training can be  accelerated and may require fewer training samples \cite{jia2020physics}. One way to inform the initialization to assist in model training and escaping local minima is to use an ML technique known as \textit{transfer learning}. In transfer learning, a model is \textit{pre-trained} on a related task prior to being fine-tuned with limited training data to fit the desired task. The pre-trained model serves as an informed initial state that ideally is closer to the desired parameters for the desired task than random initialization. One way to achieve this is to use the physics-based model's simulated data to pre-train the ML model. This is similar to the common application of pre-training in computer vision, where CNNs are often pre-trained with very large image datasets before being fine-tuned on images from the task at hand \cite{tajbakhsh2016convolutional}. 

Jia \textit{et al.} use this strategy in the context of modeling lake temperature dynamics~\cite{jia2019physics,jia2020physics}. They pre-train their Physics-Guided Recurrent Neural Network (PGRNN) models for lake temperature modeling on simulated data generated from a physics-based model and fine tune the NN with little observed data. They showed that pre-training, even using data from a physical model with an incorrect set of parameters, can still significantly reduce the training data needed for a quality model. In addition, Read et al. \cite{read2019process} demonstrated that models using both physics-guided initialization and a physics-guided loss function are able to generalize better to unseen scenarios than traditional physics-based models. In this case, physics-guided initialization allows the model to have a physically-consistent starting state prior to seeing any observations. 

Another application can be seen in robotics, where images from robotics simulations have been shown to be sufficient without any real-world data for the task of object localization \cite{tobin2017domain}, while reducing data requirements by a factor of 50 for object grasping \cite{bousmalis2018using}. Then, in autonomous vehicle training, Shah et al.~\cite{shah2018airsim} showed that pre-training the driving algorithm in a simulator built on a video game physics engine can drastically lessen data needs. More generally, we see that simulation-based pre-training of applications  allows for significantly less expensive data collection than is possible with physical robots.


Physics-guided model initialization has also been employed in chemical process modeling \cite{lu2008model,lu2009process,yan2011bayesian}. Yan et al. \cite{yan2011bayesian} use Gaussian process regression for process modeling that has been transferred and adapted from a  similar task. To adapt the transferred model, they used scale-bias correcting functions optimized through maximum likelihood estimation of parameters. Furthermore, Gaussian process models come equipped with uncertainty quantification which is also informed through initialization. A similar transfer and adapt approach is seen in Lu et al. \cite{lu2008model}, but for an ensemble of NNs transferred from related tasks. In both  studies, the similarity metrics used to find similar systems are defined by considering various common process characteristics and behaviors. 

Physics-guided initialization can also be done using a self-supervised learning method, which has been widely used in computer vision and natural language processing. In the self-supervised setting, deep neural networks learn discriminative representations using pseudo labels created from pre-defined pretext tasks. These pretext tasks are designed to extract complex patterns related to our target prediction task. 
For example, the pretext task can be defined to predict intermediate physical variables that play an important role in underlying processes. This approach can make use of a physics-based model to simulate these intermediate physical variables, which can then be used to pre-train ML models by adding supervision on hidden layers. As an illustration of this approach, Jia et al.~\cite{jia2021physics} have shown promising results for modeling temperature and flow in river networks by using upstream water variables simulated by a physics-based PRMS-SNTemp model~\cite{theurer1984instream} to pre-train hidden variables in a graph neural network.

\subsection{Physics-Guided Design of Architecture}
\label{method:arch}


Although the physics-based loss and initialization in the previous sections help constrain the search space of ML models during training, the ML architecture is often still a black box. In particular, they do not encode physical consistency or other desired physical properties into the ML architecture. A recent research direction has been to construct new ML architectures that can make use of the specific characteristics of the problem being solved. Furthermore, incorporating physics-based guidance into architecture design has the added bonus of making the previously black box algorithm more interpretable, a desirable but typically missing feature of ML models used in physical modeling. In the following paragraphs, we discuss several contexts in which physics-guided ML architectures have been used. Much of the work in this section is focused largely on neural network architectures. The modular and flexible nature of NNs in particular makes them prime candidates for architecture modification. For example, domain knowledge can be used to specify node connections that capture physics-based dependencies among variables. We also include subsections on multi-task learning and structures of Gaussian processes to show how task interrelationships or informed prior distributions can inform ML models.
 
\paragraph{Intermediate Physical Variables}
One way to embed physical principles into NN design is to ascribe physical meaning for certain neurons in the NN. It is also possible to declare physically relevant variables explicitly. In lake temperature modeling, Daw et al. \cite{daw2019physics} incorporate a physical intermediate variable as part of a monotonicity-preserving structure in the LSTM architecture. This model produces physically consistent predictions in addition to appending a dropout layer to quantify uncertainty. Muralidlar et al.  \cite{muralidhar2020phynet} used a similar approach to insert physics-constrained variables as the intermediate variables in the convolutional neural network (CNN) architecture which achieved significant improvement over state-of-the-art physics-based models on the problem of predicting drag force on particle suspensions in moving fluids. 



An additional benefit of adding physically relevant intermediate variables in an ML architecture is that they can help extract physically meaningful hidden representation that can be interpreted by domain scientists. This is particularly valuable, as standard deep learning models are limited in their interpretability since they can only extract abstract hidden variables using highly complex connected structure. This is further exacerbated given the randomness involved in the optimization process.

Another related approach is to fix one or more weights within the NN to physically meaningful values or parameters and make them non-modifiable during training. A recent approach is seen in geophysics where researchers use NNs for the waveform inversion modeling to find subsurface parameters from seismic wave data.  In Sun et al. \cite{sun2020theory}, they assign most of the parameters within a network to mimic seismic wave propagation during forward propagation of the NN, with weights corresponding to values in known governing equations. They show this leads to more robust training in addition a more interpretable NN with meaningful intermediate variables.

\paragraph{Encoding invariances and symmetries}
In physics, there is a deep connection between symmetries and invariant quantities of a system and its dynamics. For example, Noether's Law, a common paradigm in physics, demonstrates a mapping between conserved quantities of a system and the system's symmetries (e.g. translational symmetry can be shown to correspond to the conservation of momentum within a system). Therefore, if an ML model is created that is translation-invariant, the conservation of momentum becomes more likely and the model's prediction becomes more robust and generalizable. 

State-of-the-art deep learning architectures already encode certain types of invariance. For example, RNNs encode temporal invariance and CNNs can implicitly encode spatial translation, rotation, and scale invariance. In the same way, scientific modeling tasks may require other invariances based on physical laws. In turbulence modeling and fluid dynamics, Ling et al \cite{ling2016reynolds}  define a \textit{tensor basis neural network} to embed rotational invariance into a NN for improved prediction accuracy. This solves a key problem in ML models for turbulence modeling because without rotational invariance, the model evaluated on identical flows with axes defined in other directions could yield different predictions. This work alters the NN architecture by adding a higher-order multiplicative layer that ensures the predictions lie on a rotationally invariant tensor basis. In a molecular dynamics application, Anderson et al.  \cite{anderson2019cormorant} show that a rotationally covariant NN architecture can learn the behavior and properties of complex many-body physical systems.  

In a general setting, Wang et al. \cite{wang2020incorporating} show how spatiotemporal models can be made more generalizable by incorporating symmetries into deep NNs. More specifically, they demonstrated the encoding of translational symmetries, rotational symmetries, scale invariances, and uniform motion into NNs using customized convolutional layers in CNNs that enforce desired invariance properties. They also provided theoretical guarantees of invariance properties across the different designs and showed additional to significant increases in generalization performance. 

Incorporating symmetries, by informing the structure of the solution space, also has the potential to reduce the search space of an ML algorithm. This is important in the application of discovering governing equations, where the space of mathematical terms and operators is exponentially large. Though in its infancy, physics-informed architectures for discovering governing equations are beginning to be investigated by researchers. In Section \ref{obj:govern_eq}, symbolic regression is mentioned as an approach that has shown success. However, given the massive search space of mathematical operators, analytic functions, constants, and state variables, the problem can quickly become NP-hard. Udrescu et al.  \cite{udrescu2020ai} designs a recursive multidimensional version of symbolic regression that uses a NN in conjunction with techniques from physics to narrow the search space. Their idea is to use NNs to discover hidden signs of "simplicity", such  as symmetry or separability in the training data, which enables breaking the massive search space into smaller ones with fewer variables to be determined.

In the context of molecular dynamics applications, a number of researchers  \cite{behler2007generalized,zhang2018deep} have used a NN per individual atom to calculate each atom's contribution to the total energy. Then, to ensure the energy invariance with respect to the possibility of interchanging two atoms, the structure of each NN and the values of each network's weight parameters are constrained to be identical for atoms of the same element. More recently, novel deep learning architectures have been proposed for fundamental invariances in chemistry. Schutt et al. ~\cite{schutt2017schnet} proposes continuous-filter convolutional (cfconv) layers for CNNs to allow for modeling objects with arbitrary positions such as atoms in molecules, in contrast to objects described by Cartesian-gridded data such as images. Furthermore, their architecture uses atom-wise layers that incorporate inter-atomic distances that enabled the model to respect quantum-chemical constraints such as rotationally invariant energy predictions as well as energy-conserving force predictions.
As we can see, because molecular dynamics often ascribes importance to different important geometric properties of molecules (e.g. rotation),  network architectures dealing with invariances can be effective for improving performance and robustness of ML models.

Architecture modifications incorporating symmetry are also seen extensively in dynamical systems research involving differential equations. In a pioneering work by Ruthotto et al \cite{ruthotto2018deep}, three variations of CNNs are proposed to improve classifiers for images. Each variation uses mathematical theories to guide the design of the CNN based on fundamental properties of PDEs. Multiple types of modifications are made, including adding symmetry layers to guarantee the stability  expressed by the PDEs and layers that convert inputs to kinematic eigenvalues that satisfy certain physical properties. They define a parabolic CNN inspired by anisotropic filtering, a hyperbolic CNN based on Hamiltonian systems, and a second order hyperbolic CNN. Hyperbolic CNNs were found to preserve the energy in the system as intended, which set them apart from parabolic CNNs that smooth the output data, reducing the energy. Furthermore, though solving PDEs with neural networks has traditionally focused on learning
on Euclidean spaces, recently Li et al. \protect\cite{li2020fourier} proposed a new architecture which includes "Fourier neural operators" to generalize this to functional spaces. They showed it achieves greater accuracy compared to previous ML-based solvers and also can solve entire families of PDEs instead of just one.
There is a vast amount of other work using physics-guided architecture towards solving PDEs and other PDE-related applications as well which are not included in this survey (e.g. see ICLR workshop on deep learning for differential equations (\protect\cite{iclr2020diff}))

A recent direction also relating to conserved or invariant quantities is the incorporation of the Hamiltonian operator into NNs \cite{greydanus2019hamiltonian,toth2019hamiltonian,choudhary2019physics,zhong2019symplectic}. The Hamiltonian operator in physics is the primary tool for modeling the  time evolution of systems with conserved quantities, but until recently the formalism had not been integrated with NNs. Greydanus et al. \cite{greydanus2019hamiltonian} design a NN architecture that naturally learns and respects energy conservation and other invariance laws in simple mass-spring or pendulum systems. They accomplish this through predicting the Hamiltonian of the system and re-integrating instead of predicting the state of physical systems themselves. This is taken a step further in Toth et al. \cite{toth2019hamiltonian}, where they show that not only can NNs learn the Hamiltonian, but also the abstract phase space (assumed to be known in Greydanus et al. \cite{greydanus2019hamiltonian}) to more effectively model expressive densities in similar physical systems and also extend more generally to other problems in physics. Recently, the Hamiltonian-parameterized NNs above have also been expanded into NN architectures that perform additional differential equation-based integration steps based on the derivatives approximated by the Hamiltonian network \cite{chen2019symplectic}.

\paragraph{Encoding other domain-specific physical knowledge}
Various other domain-specific physical information can be encoded into architecture that doesn't exactly correspond to known invariances but provides meaningful structure to the optimization process depending on the task at hand. This can take place in many ways, including using domain-informed convolutions for CNNs, additional domain-informed discriminators in GANs, or structures informed by the physical characteristics of the problem.  For example, Sadoughi et al. \cite{sadoughi2019physics} prepend a CNN with a Fast Fourier Transform layer and a physics-guided convolution layer based on known physical information pertaining to fault detection of rolling element bearings. A similar approach is used in Sturmfels et al. \cite{sturmfels2018domain}, but the added beginning layer instead serves to segment different areas of the brain for domain guidance in neuroimaging tasks. In the context of generative models, Xie et al.  \cite{xie2018tempogan} introduce tempoGAN, which augments a general adversarial network with an additional discriminator network along with additional loss function terms that preserve temporal coherence in the generation of physics-based simulations of fluid flow. This type of approach, though found mostly in NN models, has been extended to non-NN models in Baseman et al. \cite{baseman2018physics}, where they introduce a physics-guided Markov Random Field that encodes spatial and physical properties of computer memory devices into the corresponding probabilistic dependencies.

Fan et al. \cite{fan2020solving} define new architectures to solve the inverse problem of electrical impedance tomography, where the goal is to determine the electrical conductivity distribution of an unknown medium from electrical measurements along its boundary. They define new NN layers based on a  linear approximation of both the forward and inverse maps  relying on the nonstandard form of the wavelet decomposition \cite{beylkin1991fast}.

Architecture modifications are also seen in dynamical systems research encoding principles from differential equations. Chen et al. \cite{chen2018neural} develop a continuous depth NN based on the Residual Network \cite{he2016deep} for solving ordinary differential equations. They change the traditionally discretized neuron layer depths into continuous equivalents such that hidden states can be parameterized by differential equations in continuous time. This allows for increased computational efficiency due to the simplification of the backpropagation step of training, and also creates a more scalable normalizing flow, an architectural component for solving PDEs. This is done by by parameterizing the \textit{derivative} of the hidden states of the NN as opposed to the states themselves. Then,  in a similar application, Chang et al. \cite{chang2019antisymmetricrnn} uses principles from the stability properties of differential equations in dynamical systems modeling to guide the design of  the gating mechanism and activation functions in an RNN.  

Currently, human experts have manually developed the majority of domain knowledge-encoded employed architectures, which can be a time-intensive and error-prone process. Because of this, there is increasing interest in automated neural architecture search methods  \cite{baker2016designing,elsken2018neural,hutter2019automated}. A young but promising direction in ML architecture design is to embed prior physical knowledge into neural architecture searches. Ba et al. \cite{ba2019blending} adds physically meaningful input nodes and physical operations between nodes  to the neural architecture search space to enable the search algorithm to discover more ideal physics-guided ML architectures.


\paragraph{Auxiliary Task in Multi-Task Learning}
Domain knowledge can be incorporated into ML architecture as auxiliary tasks in a multi-task learning framework. Multi-task learning allows for multiple learning tasks to be solved at the same time, ideally while exploiting commonalities and differences across tasks. This can result in improved learning efficiency and predictions for one or more of the tasks. Therefore, another way to implement physics-based learning constraints is to use an auxiliary task in a multi-task learning framework. Here, an example of an auxiliary task in a multi-task framework might be related to ensuring physically consistent solutions in addition to accurate predictions. The promise of such an approach was demonstrated for a computer vision task by  integrating auxiliary information (e.g. pose estimation) for facial landmark detection \cite{zhang2014facial}. In this paradigm, a task-constrained loss function can be formulated to allow errors of related tasks to be back-propagated jointly to improve model generalization. Early work in a computational chemistry application showed that a NN could be trained to predict energy by constructing a loss function that had penalties for both inaccuracy \textit{and} inaccurate energy derivatives with respect to time as determined by the surrounding energy force field \cite{pukrittayakamee2009simultaneous}. In particle physics, De Oliveira et al. \cite{de2017learning} uses an additional task for the discriminator network in a generative adversarial network (GAN) to satisfy certain properties of particle interaction for the production of jet images of particle energy. 


\paragraph{Physics-guided Gaussian process regression}

Gaussian process regression (GPR) \cite{williams2006gaussian} is a nonparametric, Bayesian approach to regression that is increasingly being used in ML applications. GPR has several benefits, including working well on small amounts of data and enabling uncertainty measurements on predictions. In GPR, first a Gaussian process prior must be assumed in the form of a mean function and a matrix-valued kernel or covariance function. One way to incorporate physical knowledge in GPR is to encode differential equations into the kernel \cite{swiler2020survey}. This is a key feature in Latent Force Models which attempt to use equations in the physical model of the system to inform the learning from data \cite{alvarez2009latent,luengo2016latent}. Alvarez et al. \cite{alvarez2009latent} draw inspiration from similar applications in bioinformatics \cite{lawrence2007modelling,gao2008gaussian}, which showed an increase in predictive ability in computational biology, motion capture, and geostatistics datasets. More recently, Glielmo et al.  \cite{glielmo2017accurate} propose a vectorial GPR that encodes physical knowledge in the matrix-valued kernel function. They show rotation and reflection symmetry of the interatomic force between atoms can be encoded in the Gaussian process with specific invariance-preserving covariant kernels. Furthermore, Raissi et al. \cite{raissi2018hiddenML} show that the covariance function can explicitly encode the underlying physical laws expressed by differential equations in order to solve PDEs and learn with smaller datasets.

\subsection{Hybrid Physics-ML Models}
\label{method:hybrid}

Contrary to previous sections where the focus has been on augmenting ML models specifically, numerous approaches combine physics-based models with ML models where both are operating simultaneously. We call these Hybrid Physics-ML models. In the context of Figure \protect\ref{fig:problem_diagram}, hybrid models can be viewed as replacing mechanistic model $F()$ with a new model in which $F()$ and an ML model are working together, or a subcomponent of $F()$ is replaced with ML. Hence, such methods are also referred to as ML-enhanced physical models by some researchers 
\protect\cite{ganguly2014toward}.

\subsubsection{Residual modeling}
\begin{figure}[h!]
	\centering
	\includegraphics[width=0.8\linewidth]{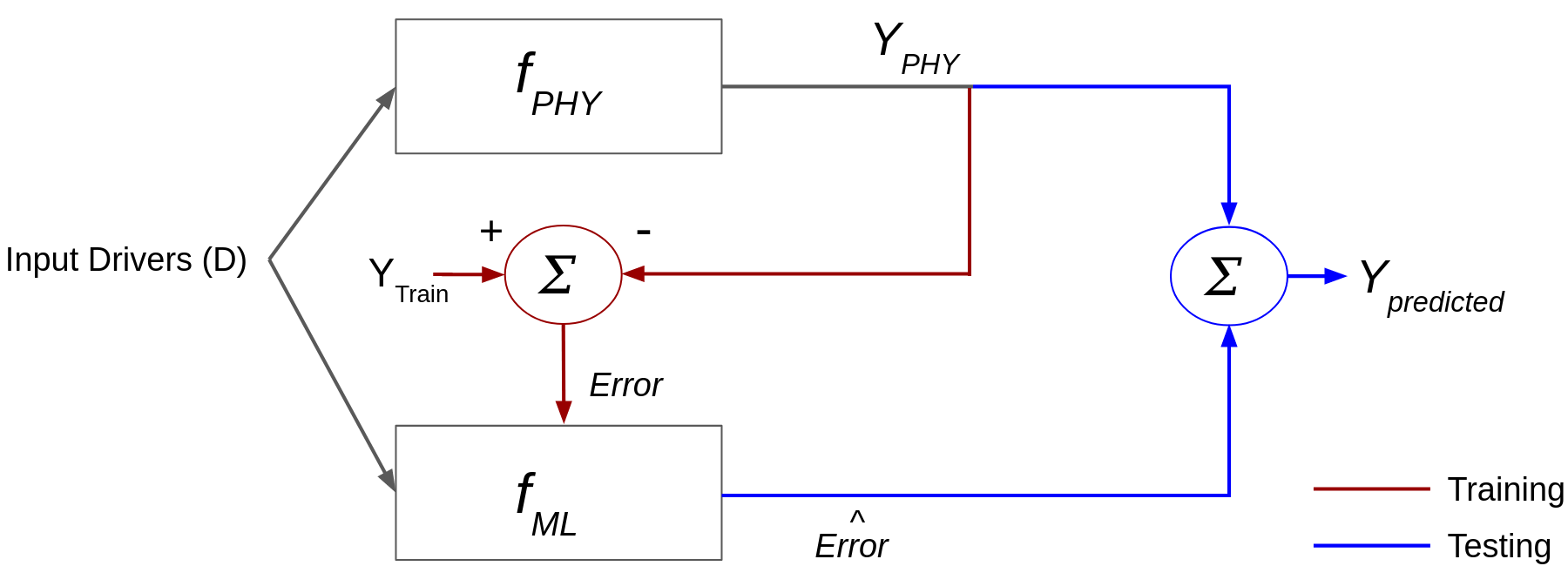}
	\caption{An illustration of the concept of residual modeling where an ML model $f_{ML}$ is trained to model the error made by the physics-based model $f_{PHY}$. Final predictions are then the sum of the predictions made by $f_{PHY}$ and the residual modeled by $f_{ML}$. Processes shown in red and blue are training and testing respectively. Figure adapted from \protect\cite{forssell1997combining}.}
	\label{fig:Residual_Modeling}
\end{figure}

The oldest and most common approach for directly addressing the imperfection of physics-based models in the scientific community is residual modeling, where an ML model (usually linear regression) learns to predict the errors, or residuals, made by a physics-based model \protect\cite{thompson1994modeling,forssell1997combining}. A visualization is shown in Figure \protect\ref{fig:Residual_Modeling}. The key concept is to learn biases of the physical model (relative to observations) and use it to correct the physical model's predictions. However, one key limitation in residual modeling is its inability to enforce physics-based constraints (like in Section \protect\ref{method:loss_func}) because such approaches model the errors instead of the physical quantities in scientific problems. 

Recently, a key area in which residual modeling has been applied is in reduced order models (ROMs) of dynamical systems (described in Section \ref{obj:roms}). After reducing model complexity to create a ROM, an ML model can be used to model the residual due to the truncation.  ROM methods were created in response to the problem of many detailed simulations being too expensive to be used in various engineering tasks including design optimization and real-time decision support. In San et al.  \cite{san2018machine,san2018neural}, a simple NN used to model the error due to the model reduction is shown to sharply reduce high error regions when applied to known differential equations. Also, in Wan et al. \cite{wan2018data}, an RNN is used to model the residual between a ROM for prediction of extreme weather events and the available data projected to a reduced-order space. 

As another example, in Kani et al. \cite{kani2017dr} a physics-driven "deep residual recurrent neural network (DR-RNN)" is proposed to find the residual minimiser of numerically discretized PDEs. Their architecture involves a stacked RNN embedded with the dynamical structure of the PDEs such that each layer of the RNN solves a layer of the residual equations. They showed that DR-RNN sharply reduces both computational cost and time discretization error of the reduced order modeling framework. Finally in Blakseth et al. \cite{blakseth2022deep}, a feed-forward neural network is used generate a corrective source term that augments the discretized governing equation of a physics-based model for improved prediction performance. This is a more advanced residual model since the ML model is modifying the governing equation itself instead of just the output.

\subsubsection{Output of physical model as input to ML model}
\begin{figure}[h!]
	\centering
	\includegraphics[width=.75\linewidth]{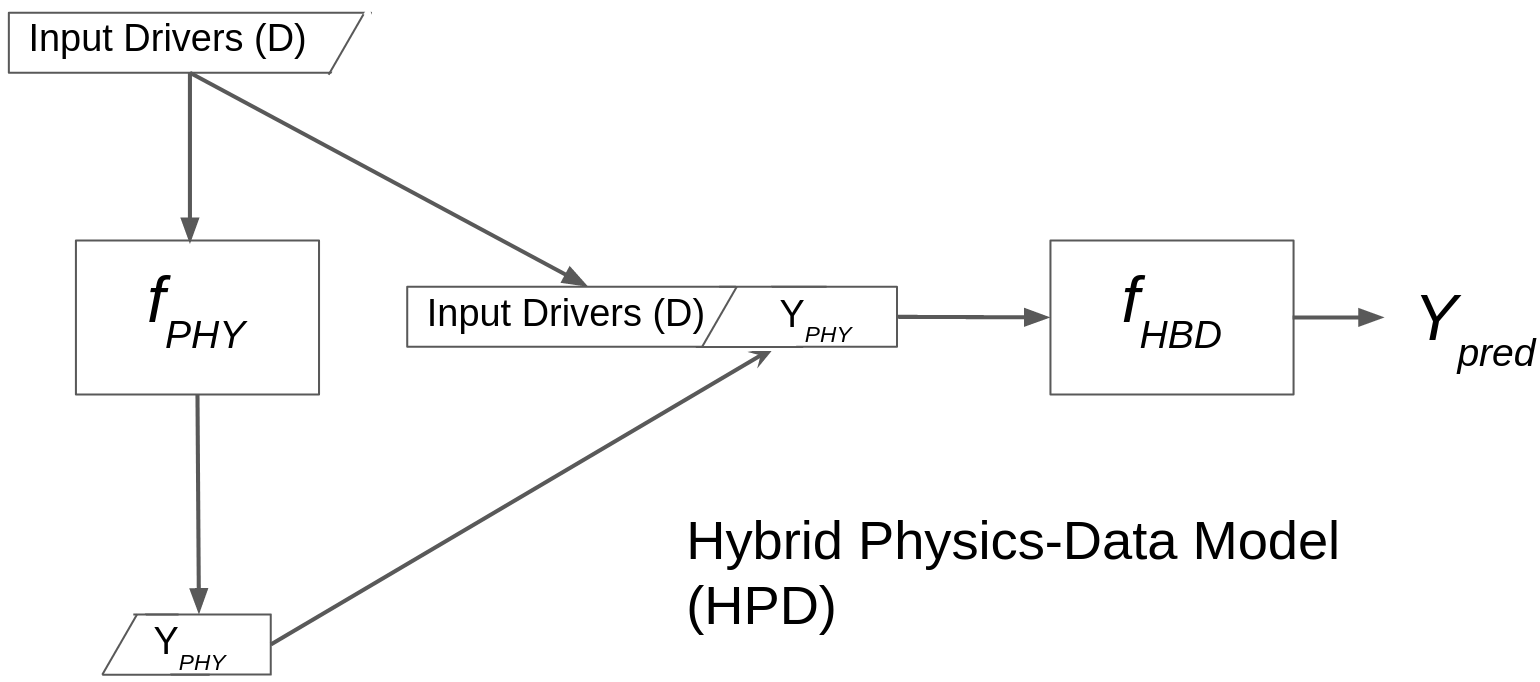}
	\caption{Diagram of a hybrid physics-ML model which accepts the output of a physical model as input to an ML model (Figure adapted from Karpatne et al.  \protect\cite{karpatne2017physics}). In the diagram, the physics-based model converts the input drivers $D$ to simulated outputs $Y_{PHY}$. Then, the hybrid physics-ML model $f_{HPD}$ jointly uses the input drivers $D$ and the simulated outputs $Y_{PHY}$ to make the final prediction $Y_{pred}$}
	\label{fig:karpatne_hybrid}
	\vspace{-.1in}
\end{figure}

In recent years many other hybrid physics-ML models have been created that extend beyond residual modeling. Another straightforward method to combine physics-based and ML models is to feed the output of a physics-based model as input to an ML model. Karpatne et al ~\cite{karpatne2017physics} showed that using the output of a physics-based model as one feature in an ML model along with inputs used to drive the physics-based model for lake temperature modeling can improve predictions. Visualization of this method is shown in Figure \ref{fig:karpatne_hybrid}. As we discuss below, there are multiple other ways of constructing a hybrid model, including replacing part of a larger physical model, or weighting predictions from different modeling paradigms depending on context.

\subsubsection{Replacing part of a physical model with ML}
In one variant of hybrid physics-ML models, ML models are used to replace one or more components of a physics-based model or to predict an intermediate quantity that is poorly modeled using physics. For example, to improve predictions of the discrepancy of Reynolds-Averaged Navier–Stokes (RANS) solvers in fluid dynamics, Parish et al.~\cite{parish2016paradigm} propose a NN to estimate variables in the turbulence closure model to account for missing physics. They show this correction to traditional turbulence models results in convincing improvements of predictions. In Hamilton et al. ~\cite{hamilton2017hybrid}, a subset of the mechanistic model's equations are replaced with data-driven nonparametric methods to improve prediction beyond the baseline process model. In Zhang et al. ~\cite{zhang2018real}, a physics-based architecture for power system state estimation embeds a deep learning model in place of traditional predicting and optimization techniques. To do this, they substitute NN layers into an unrolled version of an existing solution framework which drastically reduced the overall computational cost due to the fast forward evaluation property of NNs, but kept information of the underlying physical models of power grids and of physical constraints.

ML models for parameterization (see Section \ref{obj:parameterization}) can also be viewed as a type of hybrid modeling. The vast majority of these efforts use black box ML models   \cite{goldstein2014data,chan2017parametrization,o2018using,rasp2018deep}, but some of these parameterization models can use more sophisticated physics-guided versions of ML as we mentioned in Section \ref{method:loss_func}. 

\subsubsection{Combining predictions from both physical model and ML model}
In another class of hybrid frameworks, the overall prediction is a combination of predictions from a physical model and an ML model, where the weights depend on prediction circumstance. For example, long-range interactions (e.g. gravity) can often be more easily modeled by classical physics equations  than more stochastic short-range interactions (quantum mechanics) that are better modeled using data-driven alternatives.  Hybrid frameworks like this have been used to adaptively combine ML predictions for short-range processes and physics model predictions for long-range processes for applications in chemical reactivity~\cite{yao2018chem} and seismic activity prediction~\cite{paolucci2018broadband}. Estimator quality at a given time and location can also be used to determine whether a prediction comes from the physical model or the ML model, which was shown in Chen et al. ~\cite{chen2018pga} for air pollution estimation and in Vlachas et al. ~\cite{vlachas2018data} for dynamical system forecasting more generally. In the context of solving PDEs, Malek et al \cite{malek2006numerical} showcases a hybrid NN and traditional optimization technique to find the closed analytical form of the solution of a PDE. In this hybrid solver, there exist two terms, a term described by the NN and a term described by traditional optimization techniques. 

\subsubsection{ML informing or augmenting physics-model for inverse modeling}
Moreover, in inverse modeling, there is a growing use of hybrid models that first use physics-based models to perform the direct inversion, then use deep learning models to further refine the inverse model's predictions. Multiple works have shown an effective application for this in computed tomography (CT) reconstructions ~\cite{jin2017deep,bubba2019learning}.  Another common technique in inverse modeling of images (e.g. medical imaging, particle physics imaging), is the use of CNNs as \textit{deep image priors} \cite{ulyanov2018deep}. To simultaneously exploit data and prior knowledge, Senouf et al. ~\cite{senouf2019self} embed a CNN that serves as the image prior for a physics-based forward model for MRIs.

\subsection{Requirements and Benefits from Different Physics-ML Methodologies}
\begin{table}[h]
    \centering
    \scalebox{0.88}{
    \begin{tabular}{c|c|c|}
          {\rotatebox[origin=c]{0}{\large \thead{Physics-ML Method}}} &
            {\rotatebox[origin=c]{0}{\large \thead{Requirements}}} &
            {\rotatebox[origin=c]{0}{\large \thead{Possible Benefits}}}  
            \\
         \midrule
         \thead{Loss Function\\ \\ \\} & \thead{Known physical relationship \\(e.g. physical laws, PDEs)\\ \\} & \thead{Physical consistency, \\Improved generalization, \\ Reduced observations required,\\ Improved accuracy} \\
         \midrule
         \thead{Initialization\\ \\} & \thead{Synthetic data from mechanistic model \\ available during training} & \thead{Reduced observations required\\ Improved accuracy\\} \\
         \midrule
         \thead{Architecture\\ \\ \\}& \thead{Intermediate physical variables/processes, \\ \textit{or} hard  constraints (e.g. symmetries), \\ \textit{or}  task interrelationships, \\ \textit{or} informed prior distributions\\ \\} & \thead{Interpretability, \\ Physical consistency, \\ Improved generalization, \\ Reduced solution search space, \\ Improved accuracy }\\
         \midrule
         \thead{Hybrid\\ \\ }& \thead{Operational mechanistic model\\ available during run time \\ } & \thead{Improved accuracy \\ \\ }  \\
         \bottomrule
    \end{tabular}}
	\caption{Summary of requirements and possible benefits from different physics-ML methodologies. The left column corresponds to the four types of methods described earlier in Section \ref{sect:methods}}
	\label{tab:method_compare}
\end{table}

Methodologies for integrating scientific knowledge in ML described in this section encompass the vast majority of work on this topic. Table \protect\ref{tab:method_compare} summarizes these by listing the requirements needed for different types of methods and the corresponding possible benefits.  As we can see, depending on the context of the problem or available resources, different methods can be optimal. Hybrid methods like residual modeling are the simplest case, as they require no  process-based knowledge beyond an operational mechanistic model to be used during run time. Physics-guided loss functions require additional domain expertise to determine what terms to add to the loss function, and ML cross-validation techniques are also recommended to weight the different loss function terms. Many of the foundational works on physics-guided loss functions also include open source code that could be adapted to new applications (e.g. Raissi et al. \cite{raissi2017physics}, Read et al. \cite{read2019process}, Wang et al. \cite{wang2021understanding}). For physics-guided initialization, domain expertise can be used to determine the most relevant synthetic data for the application, but the ML can remain process-agnostic. Physics-guided architecture is often the most complex approach, where both domain and ML expertise is needed, for example, to customize neural networks by establishing physically meaningful connections and nodes. Note that there can also be multiple Physics-ML method options for a given computational benefit. For example, incorporating physical consistency into ML models can be done through weak constraints as in a loss function, hard constraints through new architectures, or indirectly through physically consistent training data from a mechanistic model simulation.

Note that for a given application-centric objective, only some of these methods may be applicable. For example, hybrid methods will not be suitable for solving PDEs since the goal of reduced computational complexity cannot be reached if the existing solver is still needed to produce the output ($y_{t}$ in Figure \protect\ref{fig:problem_diagram}). Also in the case of discovering governing equations, there often isn't a known physical model to compare to for either creating a residual model or hybrid approach. Data generation applications also do not make sense for residual modeling since the purpose is to simulate a data distribution rather than improve on a physical model.

Many of the physics-ML methods can also be combined. For example, a physics-guided loss function, physics-guided architecture, and physics-guided initialization could all be applied to an ML model. We saw in Section \protect\ref{method:loss_func} that Jia et al. \protect\cite{jia2019physics} and Read et al. \protect\cite{read2019process} in particular combined physics-guided loss functions with physics-guided initialization. Also, Karpatne et al \protect\cite{karpatne2017physics} combined a physics-guided loss function with a hybrid physics-ML framework. More recently, Jia et al \protect\cite{jia2021physics} combine physics-guided initialization and physics-guided architecture.

An overall goal of physics-ML methods presented in this section is to address resource efficiency issues (i.e., the ability to solve problems with less computational resources in the context of objectives defined in Section 2) while maintaining high predictive performance, sample efficiency, and interpretability relative to traditional ML approaches. For example, physics-ML methods for solving PDEs (Section \protect\ref{obj:solve_pde}) are likely to be more computationally efficient than direct numerical approaches and more physically consistent than traditional ML approaches. As another example, for the objective of downscaling (Section \protect\ref{obj:downscaling}), physics-ML methods can be expected to provide high resolution $y_{t}$ but at much smaller computational cost than possible via traditional mechanistic models and provide much better quality output while using fewer training samples relative to traditional ML approaches. Another major utility of physics-ML methods is to reduce the overall solution search space, which has a direct impact on sample efficiency (i.e. reduced number of observations required) and the amount of computation time taken for model training.  For example, physics-ML methods for discovering governing equations can be expected to work with much fewer observations and take less computation time relative to traditional ML methods.

\section{Areas of Current Work and Possibilities for Cross-fertilization}
\label{sec:discussion}

Table \ref{table:lit_matrix} provides a systematic organization and taxonomy of the application-centric objectives and methods of existing physics-based ML applications. This table provides a convenient organization for papers discussed in this survey and other papers that could not be discussed because of space limitations. Importantly, analysis of works within our taxonomy uncovers knowledge gaps and potential crossovers of methods between disciplines that can serve as ideas for future research.

Indeed, there are a myriad of opportunities for taking ideas across applications, objectives, and methods, as well as bringing them back to the traditional ML discipline. For example, the physics-guided NN approaches developed for aquatic sciences \cite{jia2020physics,hanson2020predicting} can be used in any application where an imperfect mechanistic model is available. Raissi et al. \cite{raissi2019deep} take physics-guided loss function methods to solve PDEs and extend them to inverse modeling problems in fluid dynamics. Furthermore, Daw et al. \cite{daw2019physics} develop a monotonicity-preserving architecture for lake temperature based on previous work done using loss function terms and a hybrid physics-ML architecture \cite{karpatne2017physics}. 
As another example Jia et al. ~\cite{jia2021physics} use physics to inform the propagation of knowledge in a graph neural network. Future research along this direction may shed light on the interpretability of hidden variables in graph neural network models and also how to build dynamic graph structures based on physics. 


From Table \ref{table:lit_matrix}, it is easy to see that several boxes are rather sparse or entirely empty, many of which represent opportunities for future work. For example, ML models for parameterization are increasingly being used in domains such as climate science and weather forecasting \cite{krasnopolsky2006complex}, all of which can benefit from the integration of physical principles. Furthermore, principles from super-resolution frameworks, originally developed in the context of computer vision applications, are beginning to be applied to downscaling to create higher resolution climate predictions \cite{vandal2017deepsd}. However, most of these do not incorporate physics (e.g., through additional loss function terms or an informed design of architecture). The fact that nearly all of the other methods (columns) except for hybrid modeling are applicable to this task shows that there is tremendous scope for new exploration, where research pursued in these columns in the context of other objectives can be applied for this objective. We also see a lot of opportunities for new research on  physics-guided initialization, where, for example, an ML algorithm could be pre-trained for inverse modeling.


Not all promising research directions are covered by Table \ref{table:lit_matrix} and the previous discussion. For instance, one promising direction is to forecast future events using ML and continuously update model states by incorporating ideas from  data assimilation \cite{kalnay2003atmospheric}. An instance of this is pursued by Dua et al. \cite{dua2011artificial} to build an ML model that predicts the parameters of a physical model using past time series data as an input. Another instance of this approach is seen in epidemiology, where Magri et al. \cite{magri2020first} used a NN for data assimilation to predict a parameter vector that captures the time evolution of a COVID-19 epidemiological model.  Such approaches can benefit from the ideas in Section \ref{sect:methods} (e.g., physics-based loss, intermediate physics variables, etc.). 
Another direction is to combine scientific knowledge and machine learning to better inform human decisions on environmental or engineering systems. For example, by using anticipated water temperature, one may build new reinforcement learning algorithms to dynamically decide when and how much water to release from reservoirs to a river network \cite{jia2021graph}. Similarly, such techniques for decision making can be used for automated control in the power plant.   

\begin{table}
\caption{Table of literature classified by objective and method}
\label{table:lit_matrix}
\scalebox{0.69}{
    \begin{tabular}[t]{c|c|c|c|c|c}
        \toprule
         & \thead{Physics-Guided\\ Loss Function\\ (3.1)}&\thead{Physics-Guided \\Initialization\\ (3.2)}&\thead{Physics-Guided\\ Architecture\\ (3.3)} &
         \multicolumn{2}{c}{\thead{Hybrid Model (3.4)\\ \\}}
        \\
        \cline{5-6}
        & & & & Residual (3.4.1) & Other (3.4.2-3.4.5)\\
        \midrule
        \makecell[t]{Improve or\\replace physical\\ model (2.1)} & 
                   
            \makecell[t]{
                    \cite{thompson1994modeling}
                    \cite{alvarez2009latent} 
                    \cite{pukrittayakamee2009simultaneous}
                    {\cite{ladicky2015data}}\\
     \cite{karpatne2017physics}
     \cite{luengo2016latent} 
                    \cite{muralidhar2018incorporating}
                    \cite{doan2019physics} \\
                    \cite{erichson2019physics} 
                    \cite{jia2019physics}
                    \cite{liu2019multi}
                    \cite{muralidhar2020phynet}\\
                    \cite{read2019process}
                    \cite{zhang2019physics}
                    \cite{hanson2020predicting}
                    \cite{hu2020physics}\\
                    \cite{magri2020first} 
                    \cite{fioretto2020predicting}
                    \cite{yazdani2019systems}
                    }
                    
            & 
            \makecell[t]{\cite{lu2008model}
                         \cite{yan2011bayesian}
                         \cite{luo2015bayesian}\\
                         \cite{tobin2017domain}
                         \cite{bousmalis2018using}
                         \cite{shah2018airsim}\\
                         \cite{jia2021physics}
                         \cite{jia2019physics}\\
                         \cite{read2019process}
                         \cite{yan2011bayesian}}
            &
            \makecell[t]{
                         \cite{ling2016reynolds} 
                         \cite{cohen2018spherical} 
                         \cite{baseman2018physics} \\
                         \cite{sturmfels2018domain} 
                         \cite{anderson2019cormorant} 
                         \cite{ba2019physics} \\
                         \cite{muralidhar2018incorporating}
                         \cite{muralidhar2020phynet} 
                         \cite{park2019physics}\\
                         \cite{sadoughi2019physics} 
                         \cite{hu2020physics}\\
                         \cite{wang2020incorporating}
                         \cite{pawar2021physics}
                         \cite{liu2020multiresolution}\\
                         \cite{sharifi2019downscaling}
                         \cite{zhang2018deep}
                         \cite{zepeda2019deep}\\
                         \cite{zhang2018deepcg}
                         \cite{schutt2017schnet}
                         \cite{dourado2020physics}}
                 
            &
            \makecell[t]{\cite{thompson1994modeling}
                    \cite{xu2015data} 
                    \cite{wang2017physics}\\ 
                    \cite{san2018machine}
                    \cite{wan2018data} 
                    \cite{wu2018physics} \\
                    \cite{liu2019multi}
                    \cite{bahari2021injecting}}
            &
            \makecell[t]{\cite{goldstein2014data}
                         \cite{grover2015deep}
                         \cite{sadowski2016synergies}\\ 
                         \cite{hamilton2017hybrid}
                         \cite{karpatne2017physics}
                         \cite{solle2017between} \\
                         \cite{chen2018pga}
                         \cite{dourado2020physics}
                         \cite{long2018hybridnet}
                         \cite{paolucci2018broadband}\\
                         \cite{yao2018chem}
                         \cite{zhang2018real}
                         \cite{vlachas2018data}\\
                         \cite{yang2019evaluation}
                         \cite{greisphysics}
                         \cite{magri2020first}\\
                         \cite{chao2022fusing}
                         \cite{blakseth2022deep}}
        \\

        \midrule
        \makecell[t]{Parameterization (2.2)} & 
            \makecell[t]{\cite{zhang2018deep}  
                         \cite{beucler2019enforcing}
                         \cite{beucler2019achieving}
                         \cite{yazdani2019systems}}
            & 
            \makecell[t]{}
            & 
            
            \makecell[t]{\cite{behler2007generalized}
                        \cite{beucler2019enforcing}
                        \cite{zhang2018real}}
            &
            
            \makecell[t]{}
            &
            
            \makecell[t]{}
        \\
        \midrule
        \makecell[t]{Downscaling (2.3)} & 
            \makecell[t]{\cite{bode2021using}
                         \cite{jiang2020meshfreeflownet}}
            & 
            \makecell[t]{}
            &
            \makecell[t]{\cite{mu2020climate}
                         \cite{vandal2018quantifying}}
            &
            
            \makecell[t]{}
            &
            
            \makecell[t]{}

        \\
        \midrule
        \makecell[t]{Reduced Order \\Models (2.4)} & 
            \makecell[t]{\cite{otto2019linearly}
                    \cite{azencot2020forecasting} 
                        \cite{lee2020model}} 
            & 
            \makecell[t]{}
            &
            \makecell[t]{\cite{kani2017dr}
                         \cite{erichson2019physics}
                         \cite{pan2020physics}}
            &
            
            \makecell[t]{\cite{kani2017dr}
                         \cite{san2018machine}
                         \cite{san2018neural}\\
                         \cite{wan2018data}
                         \cite{guo2019data}
                         \cite{pan2020physics}}
            &
            
            \makecell[t]{\cite{daniel2020model}} 
    
        \\

        \midrule
        
        \makecell[t]{Solve\\PDEs (2.5)} & 
            \makecell[t]{\cite{raissi2017physics}
                         \cite{sharma2018weakly} 
                         \cite{yang2018physics} 
                         \cite{yang2018physics2}\\ 
                         \cite{de2019deep} 
                         \cite{meng2019ppinn} 
                         \cite{raissi2019physics} 
                         \cite{shah2019encoding} \\
                         \cite{zhu2019physics}
                         \cite{dwivedi2020solution} 
                         \cite{geneva2020modeling}
                         \cite{karumuri2020simulator}\\
                         \cite{wu2020enforcing}
                         \cite{peng2020accelerating}
                         \cite{meng2020composite}\\
                         \cite{bar2019learning}
                         \cite{fang2021high}
                         \cite{zobeiry2021physics}
                         \cite{krishnapriyan2021characterizing}}
            & 
            \makecell[t]{}
            & 
            \makecell[t]{\cite{chen2018neural}
                         \cite{ruthotto2018deep}
                         \cite{chang2019antisymmetricrnn} 
                         \cite{de2019deep} \\
                         \cite{mattheakis2019physical}
                         \cite{sirignano2018dgm}
                         \cite{beck2019machine}
                         \cite{chen2019symplectic}\\ 
                         \cite{khoo2019solving}
                         \cite{fan2019multiscale}
                         \cite{yang2019highly}
                         \cite{peng2020accelerating}\\
                         \cite{meng2020composite}
                         \cite{raissi2018hiddenML}
                         \cite{kashinath2020enforcing}\\
                         \cite{mohan2020embedding}
                         \cite{li2020fourier}
                         \cite{bu2021quadratic}} 
            &
            \makecell[t]{}
            &
            \makecell[t]{\cite{malek2006numerical}}

        \\
        \midrule
            
        \makecell[t]{Inverse\\modeling\\(2.6)} & 
            \makecell[t]{\cite{raissi2019deep} 
                         \cite{kahana2020obstacle}
                         }
            & 
            \makecell[t]{}
            & 
            \makecell[t]{\cite{biswas2019prestack} 
                         \cite{fan2020solving}
                         \cite{sun2020theory}\\
                         \cite{khoo2019switchnet}
                         } 
            
            &
            \makecell[t]{{\cite{holland2019field}}} 
            &
            \makecell[t]{\cite{parish2016paradigm} 
                         \cite{hamilton2017hybrid} 
                         \cite{jin2017deep} 
                         \cite{svendsen2017joint}\\
                         \cite{camps2018physics}
                         \cite{bubba2019learning} 
                         \cite{downtonuse}
                         \cite{senouf2019self}}

        \\
        \midrule
        \makecell[t]{Discover Governing\\Equations (2.7)} & 
                         
            \makecell[t]{\cite{raissi2017physics2}
                        \cite{loiseau2018constrained} 
                        }
            & 
            \makecell[t]{}
            & 
            \makecell[t]{\cite{udrescu2020ai}
                          \cite{liu2021machine}
                          \cite{chen2020deep}\\
                          \cite{li2021physical}}
            &
            \makecell[t]{}
            &
            \makecell[t]{}
        \\

        \midrule
        \makecell[t]{Data Generation\\(2.8)} & 
            \makecell[t]{\cite{de2017learning} 
                        ~\cite{cang2018improving}    
                        \cite{bode2021using}
                        \cite{zheng2019physics}\\
                         \cite{kim2019deep}
                        \cite{wu2020enforcing} 
                        \cite{zeng2019enforcing}}
            & 
            \makecell[t]{}
            & 
            \makecell[t]{\cite{chen2018b}
                         \cite{xie2018tempogan} 
                         \cite{shah2019encoding}
                         }
            &
            \makecell[t]{}
            &
            \makecell[t]{}

        \\
        \midrule
        \makecell[t]{Uncertainty\\Quantification (2.9)} & 
            \makecell[t]{\cite{wu2016physics}
                         \cite{yang2018physics2}
                         \cite{winovich2019convpde}\\
                         \cite{yang2019adversarial} 
                         \cite{zhu2019physics} 
                         \cite{geneva2020modeling}}
            & 
            \makecell[t]{\cite{yan2011bayesian}
                         \cite{luo2015bayesian}}
            & 
            \makecell[t]{\cite{daw2019physics}
                        \cite{yang2020b}
                        \cite{winovich2019convpde}\\
                        \cite{vandal2018quantifying}}
            &
            \makecell[t]{}
            &
            \makecell[t]{\cite{dong2016hybrid}}

        \\
        \bottomrule
    \end{tabular}
    }
\end{table}

\section{Concluding Remarks}
\label{sec:conclusion}

Given the current deluge of sensor data and advances in ML methods, we envision that the merging of principles from ML and physics will play an invaluable role in the future of scientific modeling to address the pressing environmental and physical modeling problems facing society. The application-centric objectives defined in Section \protect\ref{sect:objectives} span the primary communities and disciplines that have both contributed to and benefit from physics-ML integration in a significant way. We believe these categories both provide perspective on the different ways of viewing the physics-ML integration methodologies in Section \protect\ref{sect:methods} for different purposes and also allow for coverage of a variety of disciplines that have been pursuing these ideas mostly-independently in recent years. Researchers working in one of these objectives can see how their methods fit within the taxonomy and relate them to how they are being used in other objectives.  Our hope is that this survey will accelerate the cross-pollination of ideas among these diverse research communities. 

The discussion and structure provided in this survey also serve to benefit the ML community, where, for example, techniques for adding physical constraints to loss functions can be used to enforce fairness in predictive models, or realism for data generated by GANs. Further, novel architecture designs can enable new ways to incorporate prior domain information (beyond what is usually done using Bayesian frameworks) and can lead to better interpretability.

This survey focuses primarily on improving the modeling of engineering and environmental systems that are traditionally solved using mechanistic modeling. However, the general ideas discussed here for integrating scientific knowledge in ML have wider applicability, and such research is already being pursued in many other contexts. For example, there are several interesting works in system control which often involves reinforcement learning techniques (e.g., combining model predictive control with Gaussian processes in robotics \cite{andersson2015model}, informed priors for neuroscience modeling \cite{gershman2016empirical}, physics-based reward functions in computational chemistry \cite{chophysics}, and fluidic feedback control from a cylinder~\cite{koizumi2018feedback}). Other examples include identifying features of interest in the output of computational simulations of physics-based models (e.g., high-impact weather predictions \cite{mcgovern2017using}, segmentation of climate models \cite{mudigonda2017segmenting}, and tracking phenomena from climate model data  \cite{gagne2017storm,rutz2019atmospheric}).  There is also recent work on encoding domain knowledge in geometric deep learning  \cite{bronstein2017geometric,cao2020comprehensive} that is finding increasing use in  computational chemistry \cite{duvenaud2015convolutional,fout2017protein,de2018molgan},  physics \cite{battaglia2016interaction,chang2016compositional}, hydrology \cite{jia2021physics},  geostatistics \cite{appleby2020kriging,wu2020inductive}, and neuroscience \cite{ktena2017distance}. We expect that there will be a lot of potential for cross-over of ideas amongst these different efforts that will greatly expedite research in this nascent field.

\bibliographystyle{ACM-Reference-Format}
\bibliography{acmart}

\end{document}